\documentclass[journal]{IEEEtran}

\usepackage{cite}
\usepackage{amsmath,amssymb,amsfonts,mathtools}
\usepackage{graphicx}
\usepackage{booktabs}
\usepackage[most]{tcolorbox}
\usepackage{xcolor}
\usepackage{array}
\usepackage{url}
\usepackage{balance}
\usepackage{tikz}
\usetikzlibrary{arrows.meta,positioning,fit,calc,shapes.geometric,backgrounds,patterns,shapes.symbols}
\usepackage{pifont}
\usepackage{pgfplots}
\pgfplotsset{compat=1.18}
\usepgfplotslibrary{groupplots}
\usepackage[hidelinks]{hyperref}

\newcommand{\Iset}{\mathcal{I}}
\newcommand{\Mset}{\mathcal{M}}
\newcommand{\Kset}{\mathcal{K}}
\newcommand{\E}{\mathbb{E}}
\newcommand{\Prb}{\mathbb{P}}

\newcommand{\Z}{\mathbb{Z}}
\newcommand{\one}{\mathbf{1}}
\newcommand{\CAIR}{C$^2$AIR}

\definecolor{cProp}{HTML}{1F5FA8}   
\definecolor{cBase}{HTML}{D9622B}   
\definecolor{cBin}{HTML}{2E9E6B}    
\definecolor{cBad}{HTML}{B2182B}    
\definecolor{cRan}{HTML}{6E6E6E}    
\definecolor{cVis}{HTML}{4C9BE8}
\definecolor{cAsr}{HTML}{3BB273}
\definecolor{cLsev}{HTML}{7B5CD6}
\definecolor{cLthir}{HTML}{E377A6}
\pgfplotsset{
  every axis/.append style={
    font=\footnotesize,
    tick label style={font=\scriptsize},
    label style={font=\footnotesize},
    title style={font=\footnotesize,yshift=-2pt},
    legend style={font=\scriptsize,draw=black!25,fill=white,fill opacity=0.9,text opacity=1,
                  inner sep=1.5pt,row sep=-1pt,cells={anchor=west}},
    axis line style={black!70},
    tick style={black!70},
    /pgf/number format/assume math mode=true,
    /pgf/number format/1000 sep={,},
  },
}

\title{Stochastic AI Model Placement for AI-and-RAN Coexistence in 6G Networks}

\author{Alaa~M.~Altarazi,~\IEEEmembership{Student~Member,~IEEE,}
and~Mohammad~J.~Abdel-Rahman,~\IEEEmembership{Senior~Member,~IEEE}%
\thanks{A. M. Altarazi is with the Computer Science Department, Princess Sumaya University for Technology, Amman 11941, Jordan (e-mail: a.altarazi@psut.edu.jo).}%
\thanks{M. J. Abdel-Rahman is with the Data Science Department, Princess Sumaya University for Technology, Amman 11941, Jordan, and also with the Electrical and Computer Engineering Department, Virginia Tech, Blacksburg, VA 24061, USA (e-mail: m.AbdelRahman@psut.edu.jo).}}

\begin{document}
\bstctlcite{IEEEexample:BSTcontrol}
\maketitle

\begin{abstract}
This paper considers AI-service placement over geographically distributed RAN sites equipped with GPU resources partitioned into fixed, isolated multi-instance GPU (MIG) profiles. We propose C$^2$AIR, a chance-constrained optimization model that maximizes expected AI requests served while protecting uncertain RAN workloads and enforcing probabilistic capacity reliability requirements for admitted AI services. C$^2$AIR allows a service to run at multiple sites and accounts for discrete profile, memory, and GPU-slice constraints. We evaluate C$^2$AIR on a four-site deployment calibrated from NVIDIA Aerial RAN capacity figures and MLPerf Inference results, and on seven larger scenarios with up to 15 sites and 40 AI services, computing every violation and overload probability exactly from the demand distributions. C$^2$AIR keeps the RAN violation probability within its target at every hour of the day, whereas reserving for the mean RAN demand violates it in 24\% to 47\% of the intervals. The reliability condition prevents placing AI models that would be overloaded in 85\% of the intervals at an event hour, and multi-site replication serves 2.7 times more AI requests than single-site deployment. The evaluation also shows that coarse MIG instances and a heavy demand tail can make strict RAN protection expensive. 
\end{abstract}


%
\section{Introduction}
\label{sec:introduction}

\IEEEPARstart{A}{I-and-RAN} coexistence is an emerging 6G architecture in which AI inference workloads
share accelerated computing infrastructure with conventional radio access network (RAN)
processing~\cite{Kundu2025AIRAN}. Operators that move baseband processing onto GPUs at
cell sites dimension these GPUs for the peak RAN load, so a large part of their capacity
is idle outside the busy hours. Hosting AI inference services on the same GPUs, such as
video analytics, speech recognition, and large language models (LLMs), turns this idle
capacity into revenue and brings inference close to the users. This sharing creates an
orchestration challenge: AI placement should exploit spare capacity without taking the
resources that the uncertain RAN workload needs.

The two workloads have very different requirements. RAN processing must finish every
slot within a hard real-time deadline, and a late slot is lost. AI requests tolerate
longer latency but arrive in bursts. Both demands are random and vary over the day, and
co-located AI workloads can degrade RAN processing when the GPU is shared without
protection~\cite{LoSchiavo2024YinYangRAN}. A reservation sized for the
mean RAN demand leaves the RAN short of compute whenever the demand rises above its mean,
whereas a reservation sized for the peak wastes capacity that AI could use for most of
the day. Multi-instance GPU (MIG) partitioning gives each workload an isolated share of
compute and memory~\cite{NVIDIA_MIG}, which removes interference between them. It also
makes the allocation discrete: the RAN receives a whole number of GPU slices, and each AI
service must run on one of a few fixed profile sizes. Existing AI-RAN orchestration
relies on learning-based or runtime control~\cite{Shah2025CAORA,Li2026HAF,Yao2026Beaver}
and does not give an explicit probabilistic guarantee for the RAN.

This paper studies AI-service placement across geographically distributed RAN sites
whose GPUs use fixed, isolated MIG profiles during each orchestration
interval. Before workload demand is realized, a central orchestrator selects the placement
and MIG profile of each admitted AI model and reserves GPU capacity for RAN
processing. We call the resulting chance-constrained AI-and-RAN placement model \CAIR{}.
\CAIR{} protects the RAN at every site with a chance constraint, admits an AI model only on
a profile that covers its local demand with a prescribed probability, and maximizes the
expected number of AI requests served within the slice and memory budgets of each site.
Because every probabilistic requirement involves the demand of a single site, it reduces
to a threshold computed before the solve, and \CAIR{} remains a mixed-integer linear program
that solves in well under a second.

\subsection*{Main Contributions}
The specific contributions of this paper are:
\begin{itemize}
  \item We formulate \CAIR{}, a multi-site AI model placement problem for distributed
  AI-RAN sites with fixed, isolated MIG profiles, allowing a service to be replicated
  across sites with one profile per site.
  \item We incorporate a probabilistic RAN-protection requirement that reserves enough
  compute capacity to cover uncertain RAN demand at each site with a prescribed
  reliability level.
  \item We admit only MIG profiles whose capacity covers the local AI demand with a
  prescribed probability and maximize expected served requests, while respecting
  GPU-slice and memory budgets at each site.
  \item We derive deterministic equivalents for binomial and lognormal RAN demand and for
  Gamma AI demand, which keep \CAIR{} a mixed-integer linear program, together with a
  level encoding that admits any profiled, nonlinear mapping from GPU slices to RAN
  compute.
  \item We evaluate \CAIR{} on a four-site deployment calibrated from NVIDIA Aerial RAN
  capacity figures and MLPerf Inference results, and on seven larger scenarios with up to
  15 sites, 40 AI services, and Hopper-class GPUs. All violation and overload probabilities
  are computed exactly from the demand distributions. The evaluation quantifies the effect
  of the RAN reservation policy, the AI reliability condition, replication, the reliability
  levels, the GPU budgets, the demand variability and its tail, and unplanned demand
  growth, and shows how the MIG instance sizes shape the cost of RAN protection.
\end{itemize}

\subsection*{Paper Organization}
The rest of the paper is organized as follows. Section~\ref{sec:related_work} reviews related work and positions \CAIR{} in a taxonomy of the literature. Section~\ref{sec:system_model} presents the system model, the RAN and AI demand distributions, and the problem statement. Section~\ref{sec:formulation} formulates \CAIR{} and derives its deterministic equivalent. Section~\ref{sec:evaluation} describes the evaluation setup, and Section~\ref{sec:results} reports the results. Finally, Section~\ref{sec:conclusion} concludes the paper.

\textit{Notation}: We use calligraphic letters for sets, uppercase letters for their sizes, and lowercase letters for indices.
%
\section{Related Work}
\label{sec:related_work}

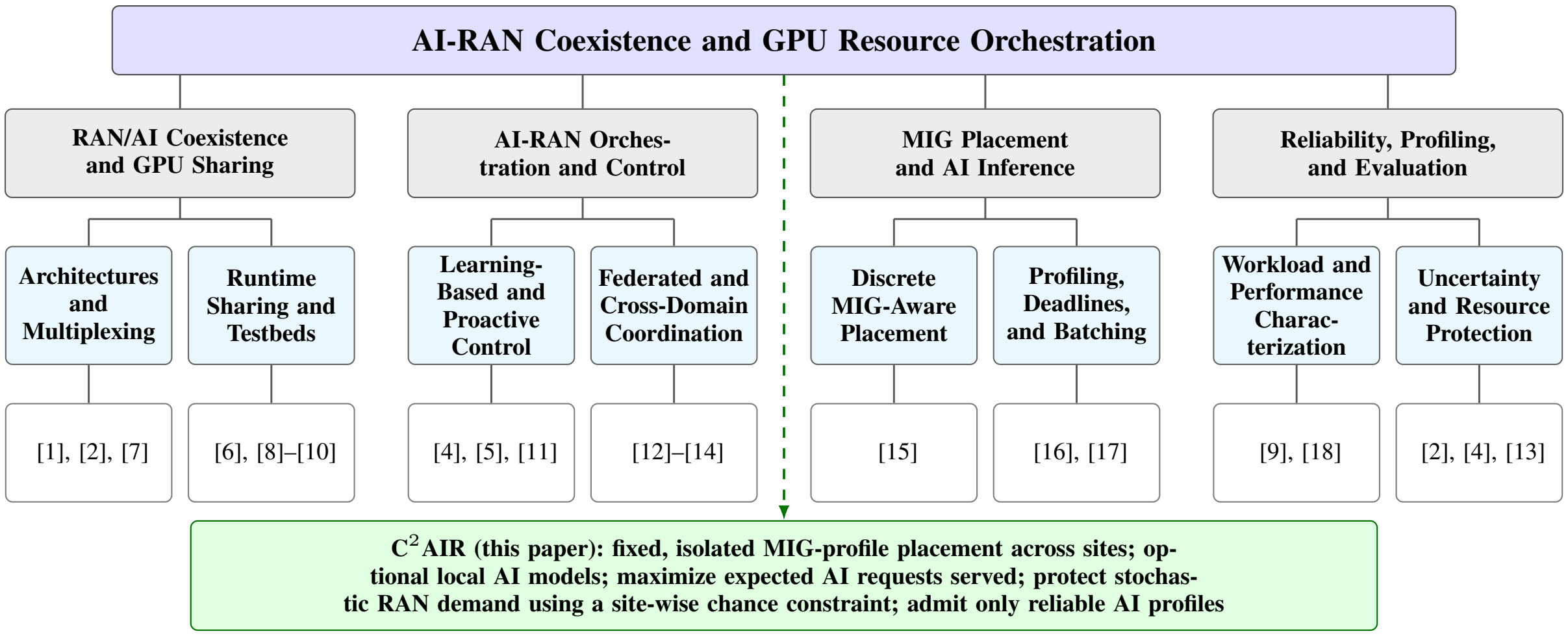
\begin{figure*}[t]
\centering
\begin{tikzpicture}[
  font=\sffamily,
  root/.style={draw=black!70, rounded corners=2pt, fill=blue!12,
    align=center, font=\small\bfseries, text width=13.4cm,
    minimum height=7mm, inner sep=4pt},
  branch/.style={draw=black!65, rounded corners=2pt, fill=gray!15,
    align=center, font=\scriptsize\bfseries, text width=3.35cm,
    minimum height=9mm, inner sep=3pt},
  subbranch/.style={draw=black!60, rounded corners=2pt, fill=cyan!7,
    align=center, font=\scriptsize\bfseries, text width=1.55cm,
    minimum height=12mm, inner sep=2pt},
  refs/.style={draw=black!50, rounded corners=2pt, fill=white,
    align=center, font=\scriptsize, text width=1.55cm,
    minimum height=10mm, inner sep=2pt},
  proposed/.style={draw=green!45!black, rounded corners=2pt, fill=green!12,
    align=center, font=\scriptsize\bfseries, text width=11.8cm,
    minimum height=10mm, inner sep=4pt},
  edge/.style={draw=black!65, line width=.55pt},
  noveltyedge/.style={draw=green!45!black, dashed, line width=.7pt, -{Latex[length=1.5mm]}}
]
\node[root] (root) at (0,0) {AI-RAN Coexistence and GPU Resource Orchestration};
\node[branch] (b1) at (-6.15,-1.15) {RAN/AI Coexistence and GPU Sharing};
\node[branch] (b2) at (-2.05,-1.15) {AI-RAN Orchestration and Control};
\node[branch] (b3) at (2.05,-1.15) {MIG Placement and AI Inference};
\node[branch] (b4) at (6.15,-1.15) {Reliability, Profiling, and Evaluation};
\draw[edge] ([xshift=-6.15cm]root.south) -- (b1.north);
\draw[edge] ([xshift=-2.05cm]root.south) -- (b2.north);
\draw[edge] ([xshift=2.05cm]root.south) -- (b3.north);
\draw[edge] ([xshift=6.15cm]root.south) -- (b4.north);

\coordinate (bus1) at (-6.15,-1.82);
\draw[edge] (b1.south) -- (bus1);
\draw[edge] (-7.08,-1.82) -- (-5.22,-1.82);
\node[subbranch] (s11) at (-7.08,-2.7) {Architectures and Multiplexing};
\node[refs] (r11) at (-7.08,-4.2) {\cite{Kundu2025AIRAN,Li2024AIRANArchitecture,LoSchiavo2024YinYangRAN}};
\node[subbranch] (s12) at (-5.22,-2.7) {Runtime Sharing and Testbeds};
\node[refs] (r12) at (-5.22,-4.2) {\cite{Kholmatov2025AoRA,ICUFN2026Testbed,Wu2026Eunoia,Yao2026Beaver}};
\draw[edge] (-7.08,-1.82) -- (s11.north);
\draw[edge] (-5.22,-1.82) -- (s12.north);
\draw[edge] (s11.south) -- (r11.north);
\draw[edge] (s12.south) -- (r12.north);

\coordinate (bus2) at (-2.05,-1.82);
\draw[edge] (b2.south) -- (bus2);
\draw[edge] (-2.98,-1.82) -- (-1.12,-1.82);
\node[subbranch] (s21) at (-2.98,-2.7) {Learning-Based and Proactive Control};
\node[refs] (r21) at (-2.98,-4.2) {\cite{Shah2025Interplay,Shah2025CAORA,Li2026HAF}};
\node[subbranch] (s22) at (-1.12,-2.7) {Federated and Cross-Domain Coordination};
\node[refs] (r22) at (-1.12,-4.2) {\cite{Shah2026FederatedCAORA,Wen2026CrossDomain,Tang2025AIServiceRepo}};
\draw[edge] (-2.98,-1.82) -- (s21.north);
\draw[edge] (-1.12,-1.82) -- (s22.north);
\draw[edge] (s21.south) -- (r21.north);
\draw[edge] (s22.south) -- (r22.north);

\coordinate (bus3) at (2.05,-1.82);
\draw[edge] (b3.south) -- (bus3);
\draw[edge] (1.12,-1.82) -- (2.98,-1.82);
\node[subbranch] (s31) at (1.12,-2.7) {Discrete MIG-Aware Placement};
\node[refs] (r31) at (1.12,-4.2) {\cite{Turkkan2024MIG}};
\node[subbranch] (s32) at (2.98,-2.7) {Profiling, Deadlines, and Batching};
\node[refs] (r32) at (2.98,-4.2) {\cite{NeuRO2026,Shen2025Batching}};
\draw[edge] (1.12,-1.82) -- (s31.north);
\draw[edge] (2.98,-1.82) -- (s32.north);
\draw[edge] (s31.south) -- (r31.north);
\draw[edge] (s32.south) -- (r32.north);

\coordinate (bus4) at (6.15,-1.82);
\draw[edge] (b4.south) -- (bus4);
\draw[edge] (5.22,-1.82) -- (7.08,-1.82);
\node[subbranch] (s41) at (5.22,-2.7) {Workload and Performance Characterization};
\node[refs] (r41) at (5.22,-4.2) {\cite{ICUFN2026Testbed,Feng2026AIRANPathway}};
\node[subbranch] (s42) at (7.08,-2.7) {Uncertainty and Resource Protection};
\node[refs] (r42) at (7.08,-4.2) {\cite{LoSchiavo2024YinYangRAN,Shah2025CAORA,Wen2026CrossDomain}};
\draw[edge] (5.22,-1.82) -- (s41.north);
\draw[edge] (7.08,-1.82) -- (s42.north);
\draw[edge] (s41.south) -- (r41.north);
\draw[edge] (s42.south) -- (r42.north);

\node[proposed] (ours) at (0,-5.45) {\CAIR{} (this paper): fixed, isolated MIG-profile placement across sites; optional local AI models; maximize expected AI requests served; protect stochastic RAN demand using a site-wise chance constraint; admit only reliable AI profiles};
\draw[noveltyedge] (root.south) -- ++(0,-0.22) -| (ours.north);
\end{tikzpicture}
\caption{Taxonomy of research related to AI-RAN coexistence and GPU resource orchestration. Published works are represented by citations only. The highlighted final node marks the position of \CAIR{}.}
\label{fig:taxonomy_ai_ran}
\end{figure*}

Sharing accelerated infrastructure between RAN processing and AI inference has been studied from several neighboring angles rather than as one mature body of work: systems that multiplex the two workloads on the same GPU, AI-RAN architectures and testbeds, orchestration frameworks that allocate resources between the workloads over time, and placement and scheduling methods for MIG-partitioned GPUs. Fig.~\ref{fig:taxonomy_ai_ran} organizes the directly relevant studies by their primary technical contribution. The branches are not mutually exclusive, so a work that contributes to more than one dimension appears more than once. This section reviews each branch and then positions \CAIR{}.

\subsection{GPU-Accelerated RAN Multiplexing and AI-RAN Architectures}
Kundu \emph{et al.} describe the AI-RAN infrastructure vision and its AI-for-RAN, AI-on-RAN, and AI-and-RAN dimensions~\cite{Kundu2025AIRAN}. Li \emph{et al.} discuss how the O-RAN architecture may evolve toward the integration of AI and communication~\cite{Li2024AIRANArchitecture}, and Feng \emph{et al.} survey the architecture, orchestration, AI-for-RAN, and AI-on-RAN dimensions of AI-RAN~\cite{Feng2026AIRANPathway}. Lo Schiavo \emph{et al.} multiplex GPU-based RAN processing with concurrent machine-learning workloads and propose a measurement-informed method that protects the RAN pipeline while preserving ML throughput~\cite{LoSchiavo2024YinYangRAN}. Kholmatov \emph{et al.} demonstrate an AI-on-RAN framework that uses the compute headroom of base stations for low-latency inference~\cite{Kholmatov2025AoRA}. An experimental testbed measures the effect of co-located LLM inference on RAN throughput and block-error performance~\cite{ICUFN2026Testbed}. Wu \emph{et al.} schedule AI and RAN GPU execution at RAN-slot granularity with Eunoia~\cite{Wu2026Eunoia}. Yao \emph{et al.} coordinate GPU compute and memory resources with Beaver to protect latency-critical vRAN processing~\cite{Yao2026Beaver}. These studies establish why the RAN must be protected when AI shares its accelerators, but none of them provides a chance-constrained placement over fixed MIG profiles.

\subsection{AI-RAN Orchestration and Control}
Shah \emph{et al.} allocate resources between RAN and AI workloads with a Soft Actor-Critic agent~\cite{Shah2025Interplay}. In their CAORA framework, Shah \emph{et al.} add O-RAN monitoring, workload forecasting, anomaly detection, and reinforcement-learning-based proactive orchestration with dynamic MIG partitioning~\cite{Shah2025CAORA}, and they extend it to multi-operator settings through federated edge orchestration~\cite{Shah2026FederatedCAORA}. Li \emph{et al.} combine slower-timescale service placement with faster-timescale deadline-aware resource sharing through hierarchical agents~\cite{Li2026HAF}. Tang \emph{et al.} study the packaging, deployment, and runtime profiling of AI services for on-demand orchestration~\cite{Tang2025AIServiceRepo}. Wen \emph{et al.} configure resources for non-collaborative cross-domain slicing using KPI prediction~\cite{Wen2026CrossDomain}. These approaches rely on learning-based control, prediction, or multi-operator coordination, whereas \CAIR{} expresses RAN protection directly as a chance constraint over fixed MIG profiles in a single orchestration interval. This is a difference in operating model rather than a claim that fixed placement is better than dynamic orchestration; the two can be combined, as discussed in Section~\ref{sec:conclusion}.

\subsection{MIG-Aware Placement and Inference Scheduling}
Turkkan \emph{et al.} optimize LLM placement on MIG GPUs without RAN workloads, accounting for the discrete MIG resources and the resulting wastage~\cite{Turkkan2024MIG}. NeuRO uses inference-time profiling for performance-aware placement of ML applications in heterogeneous edge clouds~\cite{NeuRO2026}. Shen \emph{et al.} jointly batch requests and assign workers in AI-RAN edge inference to increase the number of requests completed within their service-level objectives~\cite{Shen2025Batching}. These works motivate profile-dependent capacities and deadline-aware service models, but they do not reserve GPU slices for an uncertain RAN demand.

\subsection{Research Gap and Positioning}
\CAIR{} combines (i) discrete selection of a fixed, isolated MIG profile for each optional AI model; (ii) replication across sites, each model serving only local demand; (iii) maximization of expected AI requests served; (iv) a site-wise chance constraint on the RAN compute demand; and (v) a reliability condition that admits only profiles whose capacity covers the local AI demand with a prescribed probability. The novelty lies in combining these choices in one placement model that remains a mixed-integer linear program; the taxonomy does not by itself establish novelty. Section~\ref{sec:results} quantifies the effect of each choice against baselines that omit it. The review covers the directly relevant primary studies we identified and is not a systematic review of all AI-RAN or edge-computing publications.
%
\section{System Model, Demand Distributions, and Problem Statement}
\label{sec:system_model}

\begin{figure*}[t]
\centering
\begin{tikzpicture}[font=\footnotesize,>=Latex,
  slice/.style={draw=black!45,line width=0.4pt,minimum height=0.8cm,inner sep=0pt,anchor=south west,align=center,font=\scriptsize,text=black},
  ran/.style={slice,fill=black!15},
  vis/.style={slice,fill=cVis!30},
  asr/.style={slice,fill=cAsr!30},
  free/.style={slice,draw=black!40,dashed,fill=white,text=black!55},
  gpu/.style={draw=black!40,rounded corners=2pt,fill=black!3,line width=0.5pt},
  site/.style={draw=black!30,rounded corners=4pt,fill=white},
  radio/.style={->,orange!80!black,line width=0.9pt},
  aireq/.style={->,cVis!90!black,line width=0.9pt,dashed},
  stepn/.style={circle,draw=teal!60!black,fill=teal!15,text=black,font=\scriptsize\bfseries,inner sep=0.8pt,minimum size=0.36cm}]
\def\w{0.62}
\newcommand{\tower}[2]{%
  \draw[black!60,line width=0.8pt] (#1-0.22,#2) -- (#1,#2+1.1) -- (#1+0.22,#2);
  \draw[black!60] (#1-0.15,#2+0.32) -- (#1+0.15,#2+0.32) (#1-0.08,#2+0.65) -- (#1+0.08,#2+0.65);
  \fill[black!60] (#1,#2+1.14) circle (0.05);
  \draw[orange!70,line width=0.7pt] (#1+0.12,#2+1.24) arc (60:-60:0.13) (#1+0.2,#2+1.34) arc (60:-60:0.25);
  \draw[orange!70,line width=0.7pt] (#1-0.12,#2+1.24) arc (120:240:0.13) (#1-0.2,#2+1.34) arc (120:240:0.25);}
\newcommand{\phone}[2]{\draw[black!60,fill=white,rounded corners=0.8pt] (#1-0.09,#2-0.15) rectangle (#1+0.09,#2+0.15); \fill[black!50] (#1,#2-0.1) circle (0.016);}
\newcommand{\cam}[2]{\draw[black!60,fill=white] (#1-0.13,#2-0.09) rectangle (#1+0.07,#2+0.09); \draw[black!60,fill=white] (#1+0.07,#2) -- (#1+0.17,#2+0.08) -- (#1+0.17,#2-0.08) -- cycle;}
\newcommand{\site}[4]{%
  \draw[site] (#1,3.05) rectangle (#1+8.1,6.0);
  \node[anchor=north west,inner sep=3pt] at (#1,6.0) {\textbf{Site #2}};
  \tower{#1+0.7}{4.05}
  \phone{#1+0.45}{3.45}\phone{#1+0.72}{3.45}\cam{#1+1.05}{3.45}
  \draw[gpu] (#1+2.0,4.35) rectangle (#1+2.1+7*\w+0.1,5.35);
  \node[ran,minimum width=#3*\w cm] (ran#4) at (#1+2.1,4.45) {RAN\\$y_{#4}{=}#3$};}
\node[draw=teal!50!black,fill=teal!6,rounded corners=6pt,align=center,inner sep=4pt] (orc) at (8.8,7.05)
  {\textbf{Central orchestrator} (start of interval $t$)\\reserves $y_i$ and places $x_{m,i,k}$};
\node[stepn] at ([xshift=-0.05cm,yshift=-0.05cm]orc.north west) {1};
\draw[teal!50!black,line width=0.8pt,dashed] (orc.south) -- (8.8,6.32);
\draw[teal!50!black,line width=0.8pt,dashed] (4.05,6.32) -- (13.55,6.32);
\draw[->,teal!50!black,line width=0.8pt,dashed] (4.05,6.32) -- (4.05,6.02);
\draw[->,teal!50!black,line width=0.8pt,dashed] (13.55,6.32) -- (13.55,6.02);
\site{0}{$1$}{3}{1}
\node[vis,minimum width=2*\w cm] (v1) at (2.1+3*\w,4.45) {Vision\\\texttt{2g}};
\node[asr,minimum width=\w cm] (s1) at (2.1+5*\w,4.45) {ASR\\\texttt{1g}};
\node[free,minimum width=\w cm] at (2.1+6*\w,4.45) {free};
\site{9.5}{$I$}{5}{I}
\node[vis,minimum width=2*\w cm] (vI) at (11.6+5*\w,4.45) {Vision\\\texttt{2g}};
\node at (8.8,4.6) {\Large$\cdots$};
\draw[radio] (0.95,5.25) to[out=25,in=160] node[above,text=orange!80!black,inner sep=1pt,pos=0.45] {$\widetilde d_{1,t}$} (ran1.north west);
\draw[aireq] (1.2,3.45) -- (v1.south |- 0,3.45) node[pos=0.5,above,text=cVis!90!black,inner sep=1.5pt] {$\widetilde\lambda_{m,1,t}$} -- (v1.south);
\draw[->,cAsr!80!black,line width=0.9pt,dashed] (1.2,3.3) -- (s1.south |- 0,3.3) -- (s1.south);
\node[stepn] at (0.25,3.95) {2};
\node[stepn] at (6.15,3.75) {3};
\draw[radio] (10.45,5.25) to[out=25,in=160] node[above,text=orange!80!black,inner sep=1pt,pos=0.45] {$\widetilde d_{I,t}$} (ranI.north west);
\draw[aireq] (10.7,3.45) -- (vI.south |- 0,3.45) node[pos=0.5,above,text=cVis!90!black,inner sep=1.5pt] {$\widetilde\lambda_{m,I,t}$} -- (vI.south);
\node[stepn] at (9.75,3.95) {2};
\node[stepn] at (15.75,3.75) {3};
\begin{scope}[shift={(0.25,0.3)}]
  \node[anchor=north west,font=\footnotesize\bfseries,inner sep=0pt] at (-0.1,2.35) {(a) RAN reservation};
  \foreach \k/\h in {0/0.04,1/0.2,2/0.45,3/0.72,4/0.9,5/0.88,6/0.7,7/0.5,8/0.32,9/0.18,10/0.1,11/0.05,12/0.02}
    {\pgfmathsetmacro{\xx}{0.2+\k*0.38}
     \ifdim\xx pt>3.6pt \fill[cBad!35] (\xx-0.14,0) rectangle (\xx+0.14,\h*1.5); \else \fill[orange!35] (\xx-0.14,0) rectangle (\xx+0.14,\h*1.5); \fi}
  \draw[->,black!55] (0,0) -- (5.1,0);
  \node[anchor=north,inner sep=2pt] at (2.55,0) {$\widetilde d_{i,t}$};
  \draw[black!80,line width=0.9pt] (3.62,0) -- (3.62,1.6) node[above,inner sep=1pt] {$\phi_i(y_i)$};
  \node[text=cBad!85!black,anchor=west] at (4.2,0.55) {$\leq\varepsilon_i$};
\end{scope}
\begin{scope}[shift={(6.25,0.3)}]
  \node[anchor=north west,font=\footnotesize\bfseries,inner sep=0pt] at (-0.1,2.35) {(b) Reliable profiles};
  \fill[cBad!20] plot[domain=2.55:5,samples=30] (\x,{1.75*(\x/0.55)^2*exp(-\x/0.55)}) -- (5,0) -- (2.55,0) -- cycle;
  \draw[cVis!90!black,line width=0.9pt] plot[domain=0:5,samples=70] (\x,{1.75*(\x/0.55)^2*exp(-\x/0.55)});
  \draw[->,black!55] (0,0) -- (5.1,0);
  \node[anchor=north,inner sep=2pt] at (2.55,0) {$\widetilde\lambda_{m,i,t}$};
  \draw[black!60,line width=0.9pt,dashed] (1.45,0) -- (1.45,1.6) node[above,inner sep=1pt] {\texttt{1g} \textcolor{cBad}{\ding{55}}};
  \draw[black!80,line width=0.9pt] (2.55,0) -- (2.55,1.6) node[above,inner sep=1pt] {\texttt{2g} \textcolor{cAsr!80!black}{\ding{51}}};
  \draw[black!80,line width=0.9pt] (3.65,0) -- (3.65,1.6) node[above,inner sep=1pt] {\texttt{3g} \textcolor{cAsr!80!black}{\ding{51}}};
  \node[text=cBad!85!black,anchor=west] at (3.75,0.55) {$\leq\delta_m$};
\end{scope}
\begin{scope}[shift={(12.25,0.3)}]
  \node[anchor=north west,font=\footnotesize\bfseries,inner sep=0pt] at (-0.1,2.35) {(c) Served requests};
  \fill[cVis!20] plot[domain=0:2.55,samples=40] (\x,{1.75*(\x/0.55)^2*exp(-\x/0.55)}) -- (2.55,0) -- (0,0) -- cycle;
  \fill[pattern=north east lines,pattern color=cBase!70] plot[domain=2.55:5,samples=30] (\x,{1.75*(\x/0.55)^2*exp(-\x/0.55)}) -- (5,0) -- (2.55,0) -- cycle;
  \draw[cVis!90!black,line width=0.9pt] plot[domain=0:5,samples=70] (\x,{1.75*(\x/0.55)^2*exp(-\x/0.55)});
  \draw[->,black!55] (0,0) -- (5.1,0);
  \node[anchor=north,inner sep=2pt] at (2.55,0) {$\widetilde\lambda_{m,i,t}$};
  \draw[black!80,line width=0.9pt] (2.55,0) -- (2.55,1.6) node[above,inner sep=1pt] {$\mu_{m,k}$};
  \node[text=cVis!90!black] at (1.15,0.42) {served};
  \draw[cBase!85!black,->] (4.3,0.6) node[right,text=cBase!85!black,inner sep=1pt] {overload} -- (3.2,0.12);
\end{scope}
\end{tikzpicture}
\caption{System model: (1) the orchestrator reserves RAN slices and places AI models on reliable MIG profiles before the demand is known, (2) each site then receives random RAN and AI demand, and (3) each AI model serves only the requests of its own site, up to its profile capacity. Panels (a) to (c) illustrate the RAN chance constraint, the reliable profiles, and the served requests.}
\label{fig:system_model}
\end{figure*}
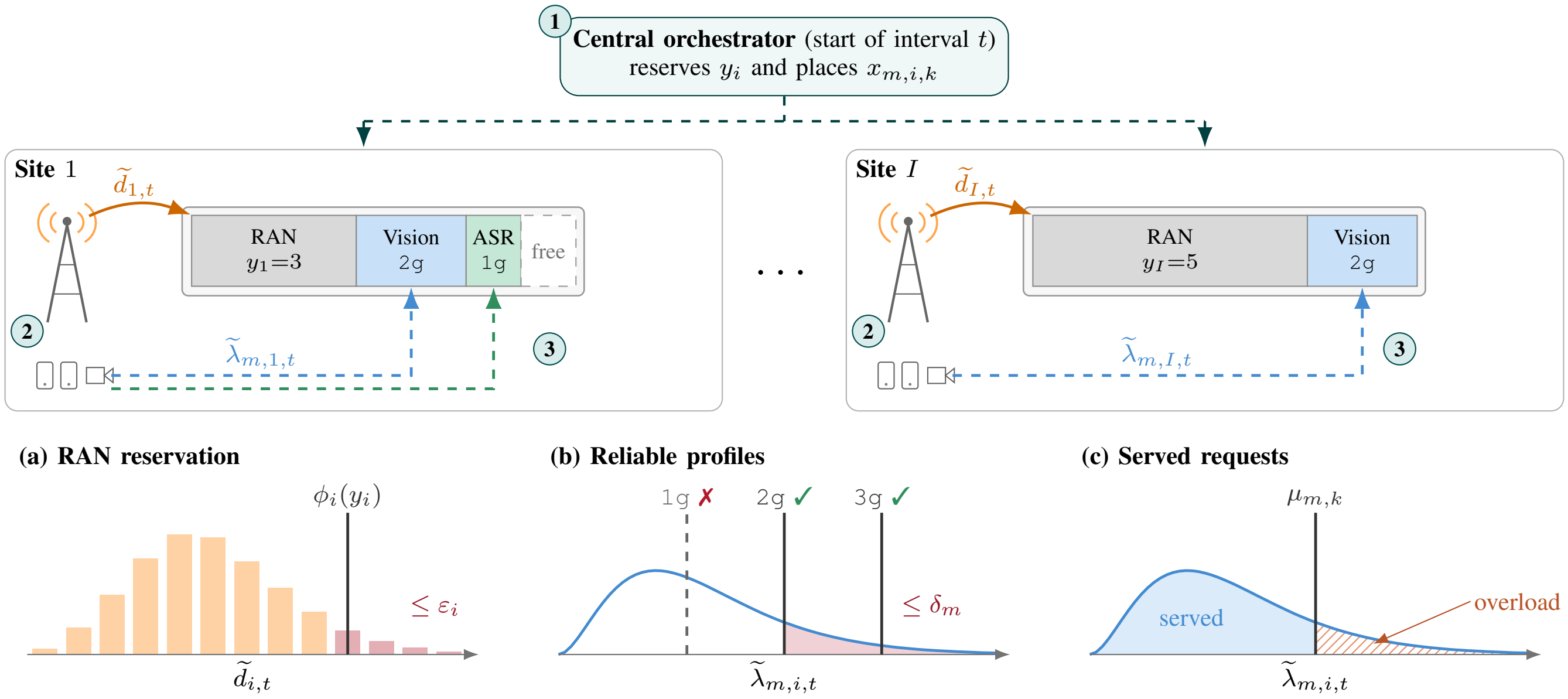

\subsection{System Model}
\label{subsec:system_model}
We consider $I$ geographically distributed RAN sites whose GPUs are shared by RAN processing and AI inference and are partitioned into fixed, isolated MIG profiles during each orchestration interval. Every site $i\in\Iset$ has $a_i$ GPU slices, $q_i$ units of GPU memory, uncertain AI request demand, and uncertain RAN compute demand, as illustrated in Fig.~\ref{fig:system_model}. A central orchestrator decides, at the beginning of each interval, which AI services to deploy, on which MIG profile each deployed AI model runs, and how many slices each site keeps for RAN processing. MIG reconfiguration within an interval is outside the scope of this paper. Table~\ref{tab:notation} summarizes the notation.

MIG profile $k\in\Kset$ consumes $s_k$ slices and provides $g_k$ units of memory, and AI service $m\in\Mset$ requires $w_m$ units of memory. Let $\widetilde{\lambda}_{m,i,t}$ and $\widetilde d_{i,t}$ denote the random demand of AI service $m$ and the random RAN compute demand at site $i$ during interval $t$. The parameter $\mu_{m,k}$ is the largest request rate that service $m$ can serve on profile $k$ while meeting its inference deadline, in requests per unit time. It is obtained offline by profiling: the service runs on profile $k$ under increasing request rates, as in the server scenario of MLPerf Inference~\cite{Reddi2020MLPerf}, and $\mu_{m,k}$ is the highest rate at which the specified latency percentile still meets the deadline. It is a measured capacity, not a decision variable.

For an AI model of service $m$ at site $i$ using profile $k$, the expected number of requests served during interval $t$, denoted by $v_{m,i,k,t}$, is
\begin{equation}
v_{m,i,k,t}=\mathbb{E}\!\left[\min\left\{\widetilde{\lambda}_{m,i,t},\mu_{m,k}\right\}\right].
\label{eq:served_requests}
\end{equation}
A profile is eligible for service $m$ at site $i$ only if it provides enough memory and satisfies the AI-capacity reliability requirement. The set of eligible profiles, denoted by $\Kset^{\mathrm{AI}}_{m,i,t}$, is
\begin{equation}
\mathcal{K}^{\mathrm{AI}}_{m,i,t}=
\big\{k\in\mathcal{K}:g_k\geq w_m,\,
\Prb\big(\widetilde{\lambda}_{m,i,t}\leq\mu_{m,k}\big)\geq1-\delta_m\big\}.
\label{eq:eligible_profiles}
\end{equation}
The probability in \eqref{eq:eligible_profiles} limits how often the local demand exceeds the profiled capacity. It guarantees the latency of the AI service only to the extent that the profiling links $\mu_{m,k}$ to the deadline, which is the case for the MLPerf server-scenario capacities used in Section~\ref{sec:evaluation}. Each AI model serves only demand originating at its own site, so models at different sites serve distinct local demand streams.

Finally, $\phi_i(\cdot)$ maps a reservation of GPU slices to the compute capacity available for RAN processing at site $i$, in the units of $\widetilde d_{i,t}$. It may be nonlinear or derived from hardware profiling; Section~\ref{subsec:det-equiv} shows how any such mapping enters \CAIR{} linearly.

\begin{table}[t]
\caption{Notation}
\label{tab:notation}
\centering
\footnotesize
\setlength{\tabcolsep}{3pt}
\renewcommand{\arraystretch}{1.08}
\begin{tabular}{>{\raggedright\arraybackslash}p{0.24\columnwidth}>{\raggedright\arraybackslash}p{0.70\columnwidth}}
\toprule
Symbol & Description\\
\midrule
\multicolumn{2}{l}{\textit{Sets and indices}}\\
$\Iset$, $i$ & RAN sites; $|\Iset|=I$\\
$\Mset$, $m$ & AI services; $|\Mset|=M$\\
$\Kset$, $k$ & MIG profiles; $|\Kset|=K$\\
$\Kset^{\mathrm{AI}}_{m,i,t}$ & Profiles eligible for service $m$ at site $i$ in interval $t$\\
$t$ & Orchestration interval\\
\midrule
\multicolumn{2}{l}{\textit{Parameters}}\\
$a_i$, $q_i$ & GPU slices and GPU memory of site $i$\\
$s_k$, $g_k$ & Slices consumed and memory provided by profile $k$\\
$w_m$ & Memory requirement of service $m$\\
$\mu_{m,k}$ & Request rate that service $m$ serves on profile $k$ within its deadline\\
$\widetilde\lambda_{m,i,t}$ & Random AI demand of service $m$ at site $i$\\
$\widetilde d_{i,t}$ & Random RAN compute demand at site $i$\\
$v_{m,i,k,t}$ & Expected requests served by service $m$ at site $i$ on profile $k$\\
$\varepsilon_i$ & Maximum RAN violation probability at site $i$\\
$\delta_m$ & Maximum overload probability of service $m$\\
$\phi_i(\cdot)$ & Mapping from reserved slices to RAN compute at site $i$\\
$n_i$, $p_i$, $r_i$ & Potential users, activity probability, and per-user RAN demand\\
$\kappa_{m,i}$, $\theta_{m,i}$ & Shape and scale of the Gamma AI demand\\
\midrule
\multicolumn{2}{l}{\textit{Decision variables}}\\
$x_{m,i,k}\in\{0,1\}$ & One if service $m$ runs at site $i$ on profile $k$\\
$y_i\in\Z_{\geq0}$ & GPU slices reserved for RAN processing at site $i$\\
\bottomrule
\end{tabular}
\end{table}

\subsection{RAN Demand Distribution}
\label{subsec:binomial}

Binomial models have been used for uncertain slice populations and for the random distribution of users across base stations. Luu \emph{et al.} use binomial slice-population models alongside uncertain per-user resource requirements~\cite{Luu2021UncertaintyAware,Luu2022AdmissionControl}. Caballero \emph{et al.} model the number of an operator's users assigned to a base station using a binomial distribution~\cite{Caballero2017SpatialLoads}. Let the number of active users at site $i$ during interval $t$ be $N_{i,t}\sim\operatorname{Binomial}(n_i,p_i)$, where $n_i$ is the number of potential users and $p_i$ is the probability that each user is active in the interval. To translate the user count into compute demand, we assume that each active user contributes a fixed amount $r_i>0$ of RAN compute demand. Then,
\begin{equation}
    \widetilde d_{i,t}=r_iN_{i,t}.
    \label{eq:binomial-demand}
\end{equation}
Since only the mean and the coefficient of variation (CV) $\mathrm{CV}_{\mathrm{RAN}}$ of the RAN demand are measured, we use the method of moments~\cite{Casella2002Inference}: matching the mean $r_in_ip_i$ and squared CV $(1-p_i)/(n_ip_i)$ of $r_iN_{i,t}$ to them gives $p_i=1/(1+n_i\,\mathrm{CV}_{\mathrm{RAN}}^2)$ and $r_i=\E[\widetilde d_{i,t}]/(n_ip_i)$. The fixed per-user demand matters: if the requirement of each user were itself random, $\widetilde d_{i,t}$ would be a random sum whose distribution must be computed before its quantile can be used.

\subsection{Lognormal RAN Demand for Comparison}
\label{subsec:lognormal}

A RAN chance constraint depends on the upper tail of the demand, which a mean and a CV do not determine. To measure how much the tail matters, we also use a lognormal model, which has positive support and a heavier right tail than the binomial model. Measurement studies have found the spatial traffic density of cellular networks to be well approximated by a lognormal distribution~\cite{Lee2014SpatialTraffic}. In this model, $\ln\widetilde d_{i,t}\sim\mathcal N(\mu_i,\sigma_i^2)$, and the method of moments gives $\sigma_i^2=\ln(1+\mathrm{CV}_{\mathrm{RAN}}^2)$ and $\mu_i=\ln\E[\widetilde d_{i,t}]-\sigma_i^2/2$. Both RAN models therefore have the same mean and CV; they differ only in the shape of the upper tail. \CAIR{} uses the binomial model throughout, and Section~\ref{subsec:res-levels} compares the two.

\subsection{AI Demand Distribution}
\label{subsec:gamma}

Model-serving studies and production LLM-serving traces describe bursty request arrivals with a Gamma distribution whose CV sets the burstiness~\cite{Li2023AlpaServe,Wang2025BurstGPT,Xiang2026ServeGen}. We model the request rate of service $m$ at site $i$ during interval $t$ as
\begin{equation}
    \widetilde\lambda_{m,i,t}\sim\mathrm{Gamma}(\kappa_{m,i},\theta_{m,i}),
    \label{eq:gamma-ai}
\end{equation}
with shape $\kappa_{m,i}$ and scale $\theta_{m,i}$, independent of the RAN demand. Since its mean is $\kappa_{m,i}\theta_{m,i}$ and its squared CV is $1/\kappa_{m,i}$, matching them to the measured mean $\E[\widetilde\lambda_{m,i,t}]$ and $\mathrm{CV}_{\mathrm{AI}}$ by the method of moments~\cite{Casella2002Inference} gives $\kappa_{m,i}=1/\mathrm{CV}_{\mathrm{AI}}^2$ and $\theta_{m,i}=\E[\widetilde\lambda_{m,i,t}]\,\mathrm{CV}_{\mathrm{AI}}^2$.

\subsection{Problem Statement}
\label{subsec:statement}
Our objective in this paper is to find \emph{the placement and MIG profile of each AI model across the RAN sites, and the number of GPU slices reserved for RAN processing at each site, that maximize the expected number of AI requests served}, where the RAN compute demand at every site must be covered with probability at least $1-\varepsilon_i$ and every AI model must run on a reliable profile within the slice and memory budgets of its site. The decisions are made at the beginning of interval $t$, before the demand is realized. A service may be replicated at multiple sites, at most one model of a service may be placed at any site, and a service may be left undeployed. The problem is single-stage: it does not model recourse after the demand is observed, cross-site request routing, or MIG reconfiguration within the interval.
%
\section{Proposed \texorpdfstring{C$^2$AIR}{C2AIR} Optimization Model}
\label{sec:formulation}

This section formulates \CAIR{}. We introduce $x_{m,i,k}$, $m\in\Mset$, $i\in\Iset$, $k\in\Kset^{\mathrm{AI}}_{m,i,t}$, as binary decision variables; $x_{m,i,k}$ equals one if service $m$ is deployed at site $i$ on MIG profile $k$, and equals zero otherwise. We also introduce $y_i\in\mathbb{Z}_{\geq0}$, $i\in\Iset$, as integer decision variables; $y_i$ is the number of GPU slices reserved for RAN processing at site $i$. Only profiles in $\Kset^{\mathrm{AI}}_{m,i,t}$ may be selected;
as defined in \eqref{eq:eligible_profiles}, a profile belongs to this set only if it provides
at least $w_m$ units of memory and its capacity $\mu_{m,k}$ covers the local AI demand with
probability at least $1-\delta_m$.

\subsection{Objective Function}
The objective is to maximize the total expected number of AI requests served across all
sites and services during interval $t$. A deployed AI model contributes its expected
served requests $v_{m,i,k,t}$ from \eqref{eq:served_requests}, so the objective is
\begin{equation}
\underset{\{x_{m,i,k},\,y_i\}}{\mathrm{maximize}} \quad
\sum_{i\in\Iset}\sum_{m\in\Mset}
\sum_{k\in\Kset^{\mathrm{AI}}_{m,i,t}}
v_{m,i,k,t} \; x_{m,i,k}.
\label{eq:objective}
\end{equation}
The objective counts the requests of all services equally. An operator that values some services more, for example LLM requests over video frames, can weight each term; the weights do not change the structure of \CAIR{}.

\subsection{Constraints}

\subsubsection{Chance Constraint for RAN Protection}
Let $\phi_i(\cdot)$ be a function that maps a reservation of $y_i$ GPU slices to the computing capacity available
for RAN processing at site $i$, in the units of $\widetilde d_{i,t}$.
At every site, the reserved RAN capacity $\phi_i(y_i)$ must
be sufficient to cover the random demand with probability at least $1-\varepsilon_i$. This requirement is expressed by the chance constraint $\Prb\big(\widetilde d_{i,t}\leq\phi_i(y_i)\big)\geq 1-\varepsilon_i, \forall i\in\Iset$. Each chance constraint involves the demand of a single site, so the constraints of different sites are separate and no joint distribution across sites is needed.

\subsubsection{GPU-Slice Capacity Constraint}
The GPU slices assigned to AI models and those reserved for RAN processing cannot
exceed the total slice capacity available at a site. Hence, the constraint $\sum_{m\in\Mset}\sum_{k\in\Kset^{\mathrm{AI}}_{m,i,t}} s_k \; x_{m,i,k}+y_i\leq a_i$ is imposed for every site $i\in\Iset$.
Since $x_{m,i,k}$ equals one only for the profile on which service $m$ runs at site $i$, the double sum adds the slices $s_k$ of all AI models placed at site $i$. Adding the $y_i$ slices reserved for the RAN gives the number of slices in use, which cannot exceed the $a_i$ slices of the site, so the RAN reservation and the AI models never share a slice.

\subsubsection{GPU-Memory Capacity Constraint}
The total memory consumed by deployed AI models must not exceed the GPU-memory budget
of the corresponding site. This requirement is represented by $\sum_{m\in\Mset}\sum_{k\in\Kset^{\mathrm{AI}}_{m,i,t}} g_k \; x_{m,i,k}\leq q_i, \forall i\in\Iset$.
In the same way, the double sum adds the memory $g_k$ of the profile of every AI model placed at site $i$, and the total cannot exceed the memory $q_i$ of the site. Because every profile in $\Kset^{\mathrm{AI}}_{m,i,t}$ provides at least $w_m$ units of memory by \eqref{eq:eligible_profiles}, each AI model also has enough memory for its weights.

\subsubsection{At Most One Profile per Service at Each Site}
A service may be replicated across different sites, but \CAIR{} allows at most
one profile and one model of a given service at any individual site. This is enforced by $\sum_{k\in\Kset^{\mathrm{AI}}_{m,i,t}}x_{m,i,k}\leq 1, \forall m\in\Mset$ and $i\in\Iset$. The local demand of a service at a site is therefore served by one AI model and is never split.

\subsection{Mathematical Formulation}
Combining the objective function and all constraints yields the following single-stage
chance-constrained mixed-integer program, which Section~\ref{subsec:det-equiv} turns into a mixed-integer linear program (MILP):
\begin{tcolorbox}[
    enhanced jigsaw,
    breakable,
    pad at break*=1mm,
    enlarge top initially by=2mm,
    attach boxed title to top center={yshift=-3mm,yshifttext=-1mm},
    colback=white,
    colframe=blue!60!black,
    colbacktitle=blue!60!black,
    title=\textsc{\CAIR{} Formulation}
]
\begin{align}
&\underset{\{x_{m,i,k},\,y_i\}}{\mathrm{maximize}}
\quad \sum_{i\in\Iset}\sum_{m\in\Mset}
\sum_{k\in\Kset^{\mathrm{AI}}_{m,i,t}}
 v_{m,i,k,t} \; x_{m,i,k}
\label{prob:objective}\\
&\text{subject to:}\nonumber\\
&\Prb\!\left(\widetilde d_{i,t}\leq\phi_i(y_i)\right)
\geq 1-\varepsilon_i, \forall i\in\Iset,
\label{prob:ran_chance}\\
&\sum_{m\in\Mset}\sum_{k\in\Kset^{\mathrm{AI}}_{m,i,t}}
 s_k \; x_{m,i,k}+y_i\leq a_i, \forall i\in\Iset,
\label{prob:slices}\\
&\sum_{m\in\Mset}\sum_{k\in\Kset^{\mathrm{AI}}_{m,i,t}}
 g_k \; x_{m,i,k}\leq q_i, \forall i\in\Iset,
\label{prob:memory}\\
&\sum_{k\in\Kset^{\mathrm{AI}}_{m,i,t}}x_{m,i,k}\leq 1, \forall m\in\Mset,\ \forall i\in\Iset,
\label{prob:one_profile}\\
&x_{m,i,k}\in\{0,1\}, \forall m\in\Mset,\ \forall i\in\Iset,
\ \forall k\in\Kset^{\mathrm{AI}}_{m,i,t},
\label{prob:binary}\\
&y_i\in\mathbb{Z}_{\geq0}, \forall i\in\Iset.
\label{prob:reservation_domain}
\end{align}
\end{tcolorbox}
The slice and memory
constraints assume that the selected profiles can be combined on the target GPU; otherwise,
hardware-specific compatibility constraints must be added, as we do for the Hopper GPUs of Section~\ref{subsec:setup-scenarios}.

\subsection{Deterministic Equivalent MILP}
\label{subsec:det-equiv}

\subsubsection{RAN Protection Chance Constraint}
\label{subsec:de-binomial}
With the binomial demand \eqref{eq:binomial-demand}, the chance constraint \eqref{prob:ran_chance} becomes
\begin{equation}
    \Prb\big(r_iN_{i,t}\leq\phi_i(y_i)\big)\geq1-\varepsilon_i,
    \qquad \forall i\in\Iset.
    \label{eq:binomial-cc}
\end{equation}
Since $N_{i,t}$ is integer valued, \eqref{eq:binomial-cc} is equivalent to $F_{\operatorname{Bin}(n_i,p_i)}\big(\lfloor\phi_i(y_i)/r_i\rfloor\big)\geq1-\varepsilon_i$, where $F_{\operatorname{Bin}(n_i,p_i)}(\cdot)$ is the binomial cumulative distribution function (CDF). The floor appears because $r_iN_{i,t}\leq\phi_i(y_i)$ is the same event as $N_{i,t}\leq\phi_i(y_i)/r_i$, and an integer $N_{i,t}$ does not exceed $\phi_i(y_i)/r_i$ exactly when it does not exceed the largest integer below it, $\lfloor\phi_i(y_i)/r_i\rfloor$. Let $Q_i^{\operatorname{Bin}}\triangleq\min\{q\in\{0,\ldots,n_i\}:F_{\operatorname{Bin}(n_i,p_i)}(q)\geq1-\varepsilon_i\}$ be the binomial quantile. Because the CDF is nondecreasing, the condition holds if and only if $\lfloor\phi_i(y_i)/r_i\rfloor\geq Q_i^{\operatorname{Bin}}$, that is, if and only if
\begin{equation}
    \phi_i(y_i)\geq r_iQ_i^{\operatorname{Bin}},
    \qquad \forall i\in\Iset.
    \label{eq:dep-binomial}
\end{equation}
For the lognormal model of Section~\ref{subsec:lognormal}, $\Prb\big(\widetilde d_{i,t}\leq\phi_i(y_i)\big)=\Phi\big((\ln\phi_i(y_i)-\mu_i)/\sigma_i\big)$ for positive capacity, where $\Phi$ is the standard normal CDF. Since $\Phi$ and the logarithm are increasing, the chance constraint becomes
\begin{equation}
    \phi_i(y_i)\geq\exp\!\big(\mu_i+\sigma_i\Phi^{-1}(1-\varepsilon_i)\big),
    \qquad \forall i\in\Iset.
    \label{eq:dep-lognormal}
\end{equation}
In both cases the right-hand side is a constant once the distribution parameters and $\varepsilon_i$ are fixed, so the chance constraint is replaced by a deterministic capacity requirement.

\subsubsection{Expected Number of AI Requests Served}
\label{subsec:de-gamma}
With the Gamma demand \eqref{eq:gamma-ai} and $(\kappa,\theta)=(\kappa_{m,i},\theta_{m,i})$, it can be shown that the reliability condition in \eqref{eq:eligible_profiles} holds if and only if
\begin{equation}
    \mu_{m,k}\geq F^{-1}_{\Gamma(\kappa,\theta)}(1-\delta_m),
    \label{eq:gamma-reliable}
\end{equation}
where $F_{\Gamma(\kappa,\theta)}$ is the Gamma CDF. The set $\Kset^{\mathrm{AI}}_{m,i,t}$ is therefore known before the solve. Using the indicator function $\one\{\cdot\}$ and $(\widetilde\lambda,\mu)=(\widetilde\lambda_{m,i,t},\mu_{m,k})$, $v_{m,i,k,t}$ in~\eqref{eq:served_requests} can be expressed as
\begin{align}
v_{m,i,k,t} & = \E\big[\widetilde\lambda\,\one\{\widetilde\lambda\leq\mu\}\big] + \E\big[\mu\,\one\{\widetilde\lambda>\mu\}\big] \nonumber \\
& = \kappa\theta F_{\Gamma(\kappa+1,\theta)}(\mu) +\mu\big[1-F_{\Gamma(\kappa,\theta)}(\mu)\big]. \label{eq:ai-served}
\end{align}
The first term counts the demand served in the intervals in which the AI model keeps up with it, and the second counts the capacity delivered in the intervals in which the model is overloaded. The reliability condition needs the quantile in \eqref{eq:gamma-reliable} and the objective needs the partial expectation in \eqref{eq:ai-served}; the mean alone determines neither. The Appendix derives \eqref{eq:dep-lognormal}, \eqref{eq:gamma-reliable}, and \eqref{eq:ai-served}.

\subsubsection{Capacity Mapping}
\label{subsec:capacity-mapping}
The mapping $\phi_i$ need not be linear. We write it with one binary variable $z_{i,l}$ per admissible reservation level $l$, so that $\phi_i(y_i)=\sum_l\phi_i(l)\,z_{i,l}$, $y_i=\sum_l l\,z_{i,l}$, and $\sum_l z_{i,l}=1$. With this encoding, \eqref{eq:dep-binomial} and \eqref{eq:dep-lognormal} are linear for any profiled mapping, and levels that the hardware does not support are simply left out of the sums.

All thresholds, eligible profiles, and served-request coefficients are computed before the solve, so \CAIR{} is an MILP that a standard solver handles directly.
%
\section{Evaluation Setup}
\label{sec:evaluation}

We evaluate \CAIR{} in two settings. The first is a four-site deployment whose parameters are calibrated from published hardware and benchmark measurements; it is small enough to inspect every placement decision. The second is a set of seven larger scenarios with up to 15 sites and 40 AI services on Hopper-class GPUs, which tests whether the observations carry over to a larger scale and to a different GPU generation. No demand data are sampled: the distribution parameters are set from the published measurements by the method of moments, and all thresholds, violation probabilities, and overload probabilities are computed exactly from the binomial, lognormal, and Gamma distributions. Because the evaluation uses the distributions assumed by \CAIR{}, its plans meet their targets by construction; the purpose of the evaluation is to measure what the baselines lose and what the protection costs. Table~\ref{tab:eval-params} collects the parameters of the four-site deployment with their sources.

\begin{table}[t]
\caption{Parameters of the Four-Site Deployment. Values Are Listed for Sites 1 to 4 Where They Differ by Site.}
\label{tab:eval-params}
\centering
\footnotesize
\setlength{\tabcolsep}{2.5pt}
\renewcommand{\arraystretch}{1.12}
\begin{tabular}{>{\raggedright\arraybackslash}p{0.28\columnwidth}>{\raggedright\arraybackslash}p{0.14\columnwidth}>{\raggedright\arraybackslash}p{0.33\columnwidth}>{\raggedright\arraybackslash}p{0.16\columnwidth}}
\toprule
Parameter & Symbol & Value & Source \\
\midrule
\multicolumn{4}{l}{\textit{GPU resources}}\\
GPU model & & A100-SXM4-80GB & \cite{NVIDIA_A100} \\
GPUs per site & & 1, 2, 2, 3 & scenario \\
MIG slices per site & $a_i$ & 7, 14, 14, 21 & \cite{NVIDIA_MIG} \\
GPU memory per site & $q_i$ & 80, 160, 160, 240 GB & \cite{NVIDIA_A100} \\
Profile slices, memory & $(s_k,g_k)$ & (1,10), (2,20), (3,40), (4,40), (7,80) & \cite{NVIDIA_MIG} \\
Capacity mapping & $\phi_i(l)$ & $14\,l$ SMs & \cite{NVIDIA_MIG} \\
\midrule
\multicolumn{4}{l}{\textit{RAN demand}}\\
Cells per site & & 4, 7, 9, 13 & scenario \\
SMs per cell & & 9 (average), 18 (peak) & \cite{NVIDIA_Aerial231} \\
Mean RAN demand & $\E[\widetilde d_{i,t}]$ & 36, 63, 81, 117 SMs & derived \\
CV of RAN demand & $\mathrm{CV}_{\mathrm{RAN}}$ & 0.33 & \cite{NVIDIA_Aerial231,Lee2014SpatialTraffic} \\
Binomial users, activity & $(n_i,p_i)$ & 100, 0.0841 & \cite{Luu2021UncertaintyAware,Caballero2017SpatialLoads} \\
\midrule
\multicolumn{4}{l}{\textit{AI services and demand}}\\
Profile capacity & $\mu_{m,k}$ & Table~\ref{tab:eval-ai} & \cite{Reddi2020MLPerf,MLPerfV30,MLPerfV31} \\
Memory requirement & $w_m$ & 4, 4, 20, 35 GB & model size \\
Mean AI demand & $\E[\widetilde\lambda_{m,i,t}]$ & Table~\ref{tab:eval-ai} & scenario \\
CV of AI demand & $\mathrm{CV}_{\mathrm{AI}}$ & 0.30 & \cite{Li2023AlpaServe,Wang2025BurstGPT,Xiang2026ServeGen} \\
\midrule
\multicolumn{4}{l}{\textit{Demand levels and targets}}\\
Time-of-day factor & & 0.11, 0.5, 1.0, 1.3 & \cite{Wang2015TrafficPatterns} \\
RAN violation level & $\varepsilon_i$ & 0.05 (0.10 to 0.001) & this work \\
AI reliability level & $\delta_m$ & 0.05 (0.20 to 0.01) & this work \\
\bottomrule
\end{tabular}
\end{table}

\subsection{GPU and MIG Resources}
\label{subsec:setup-gpu}

Sites 1 to 4 have 1, 2, 2, and 3 NVIDIA A100-SXM4-80GB GPUs~\cite{NVIDIA_A100}. MIG partitions each GPU into seven compute slices of 14 streaming multiprocessors (SMs)~\cite{NVIDIA_MIG}, so the slice budgets are $a_i=7$, 14, 14, and 21 and the memory budgets $q_i=80$, 160, 160, and 240~GB. An AI model runs on one of the five A100 profiles \texttt{1g.10gb}, \texttt{2g.20gb}, \texttt{3g.40gb}, \texttt{4g.40gb}, and \texttt{7g.80gb}, where the leading number is the slice count $s_k$ and the memory is $g_k$. The RAN reservation is counted in slices, and $l$ reserved slices provide $\phi_i(l)=14\,l$ SMs. Because the mapping enters \CAIR{} through the level variables of Section~\ref{subsec:capacity-mapping}, a measured or nonlinear mapping could replace it without changing the formulation.

\subsection{RAN Demand}
\label{subsec:setup-ran}

NVIDIA Aerial SDK~23-1 runs 12 average-load or 6 peak-load 100~MHz 4T4R cells on one A100X GPU~\cite{NVIDIA_Aerial231}, so over the 108 SMs of the GPU a cell needs about 9 SMs at average load and 18 SMs at peak load. The four sites host 4, 7, 9, and 13 cells, so their mean busy-hour RAN demands are 36, 63, 81, and 117 SMs. If every cell ran at peak load, the RAN would occupy 64\% to 83\% of the GPU capacity of its site, a usual dimensioning margin for a RAN pool.

The cell count sets the mean of the RAN demand and the CV sets its spread. We use $\mathrm{CV}_{\mathrm{RAN}}=0.33$, the value at which the 99th percentile of the lognormal model equals twice the mean, the ratio between a peak-load and an average-load cell. The binomial model uses $n_i=100$ potential users, which with the method of moments of Section~\ref{subsec:binomial} gives $p_i=0.0841$ and $r_i$ equal to the mean demand divided by $n_ip_i$. Its 99th percentile is 1.78 times the mean, against 2.01 times for the lognormal model with the same mean and CV.

\subsection{AI Services and Demand}
\label{subsec:setup-ai}

The four AI services are object detection (Vision), automatic speech recognition (ASR), and two LLMs with 7B and 13B parameters (LLM7B and LLM13B). We derive the capacities $\mu_{m,k}$ from MLPerf Inference server-scenario results on A100-SXM-80GB systems, in which 99\% of the queries meet a latency bound~\cite{Reddi2020MLPerf,MLPerfV30,MLPerfV31}. The per-GPU rates are 700.3 frames/s for RetinaNet at 100~ms, 13{,}000 utterances/s for RNN-T at 1~s, and 4.06 requests/s for GPT-J-6B at 20~s. The GPT-J rate is rescaled by parameter count to the 7B and 13B models~\cite{Touvron2023Llama2}, because token generation is limited by reading the weights from memory. A profile with $s_k$ slices receives the share $14s_k/108$ of the full-GPU rate, and $\mu_{m,k}=0$ when the model does not fit in $g_k$. The memory requirements $w_m=4$, 4, 20, and 35~GB cover the FP16 weights of the LLMs (13.5 and 26~GB) and their key-value cache.

Table~\ref{tab:eval-ai} lists the capacities and the mean busy-hour demands. The Vision demand corresponds to 8 to 20 cameras analyzed at 10 frames/s per site. The AI demand is Gamma distributed with $\mathrm{CV}_{\mathrm{AI}}=0.30$.

\begin{table}[t]
\caption{Profile Capacities $\mu_{m,k}$ (Requests/s) Derived From MLPerf Inference Results, and Mean Busy-Hour AI Demand at Sites 1 to 4. A Dash Marks a Profile With Too Little Memory for the Model.}
\label{tab:eval-ai}
\centering
\footnotesize
\setlength{\tabcolsep}{2.4pt}
\begin{tabular}{lrrrrrrrrr}
\toprule
 & \multicolumn{5}{c}{$\mu_{m,k}$} & \multicolumn{4}{c}{$\E[\widetilde\lambda_{m,i,t}]$}\\
\cmidrule(lr){2-6}\cmidrule(lr){7-10}
Service & 1g & 2g & 3g & 4g & 7g & Site 1 & Site 2 & Site 3 & Site 4\\
\midrule
Vision & 90.8 & 182 & 272 & 363 & 635 & 80 & 100 & 200 & 160\\
ASR & 1,685 & 3,370 & 5,056 & 6,741 & 11,796 & 20 & 30 & 40 & 50\\
LLM7B & -- & 0.945 & 1.42 & 1.89 & 3.31 & 0.4 & 0.5 & 0.8 & 0.7\\
LLM13B & -- & -- & 0.735 & 0.98 & 1.71 & 0.2 & 0.25 & 0.4 & 0.35\\
\bottomrule
\end{tabular}
\end{table}

\subsection{Demand Levels and Reliability Targets}
\label{subsec:setup-profiles}

\CAIR{} is solved once per orchestration interval, so we evaluate it at four demand levels of a day. All mean demands, RAN and AI, are multiplied by 0.11 (valley), 0.5 (shoulder hour), 1.0 (busy hour), or 1.3 (an event hour 30\% above the usual busy hour). The valley factor follows the peak-to-valley ratio of about 9 measured on 9{,}600 cellular towers~\cite{Wang2015TrafficPatterns}. For the binomial model, a demand level scales the activity probability $p_i$ while the per-user demand $r_i$ stays fixed, so a quieter hour has fewer active users, each with the same compute need. Unless stated otherwise, the demand is at the busy hour and $\varepsilon_i=\delta_m=0.05$; Sections~\ref{subsec:res-levels} and~\ref{subsec:res-delta} vary $\varepsilon_i$ from 0.10 to 0.001 and $\delta_m$ from 0.20 to 0.01.

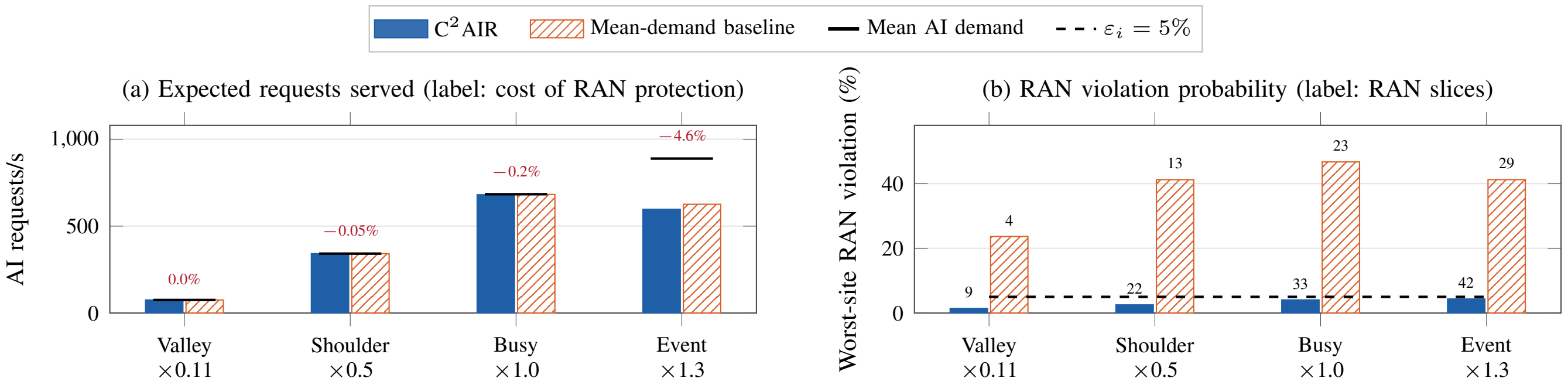
\begin{figure*}[t]
%
\centering
\ref{lg:tod}\\[1pt]
\begin{minipage}[b]{0.49\textwidth}
\centering
\begin{tikzpicture}
\begin{axis}[width=\linewidth,
  height=3.7cm,
  ybar=1pt,
  bar width=12pt,
  symbolic x coords={V,Sh,B,Su},
  xtick=data,
  xticklabels={{Valley\\$\times$0.11},{Shoulder\\$\times$0.5},{Busy\\$\times$1.0},{Event\\$\times$1.3}},
  xticklabel style={align=center},
  enlarge x limits=0.15,
  area legend,
  ymajorgrids,
  grid style={black!10},
  ymin=0,
  ymax=1080,
  ylabel={AI requests/s},
  title={(a) Expected requests served (label: cost of RAN protection)},
  legend to name=lg:tod,
  legend columns=-1,
  legend style={draw=black!25,/tikz/every even column/.append style={column sep=8pt}}]
\addplot[fill=cProp,draw=cProp] coordinates {(V,75.17) (Sh,341.53) (B,680.84) (Su,597.18)};\addlegendentry{C$^2$AIR}
\addplot[pattern=north east lines,pattern color=cBase,draw=cBase] coordinates {(V,75.17) (Sh,341.70) (B,682.25) (Su,625.78)};\addlegendentry{Mean-demand baseline}
\addplot[only marks,mark=-,mark size=10pt,thick,black,legend image code/.code={\draw[black,very thick] (0cm,0cm) -- (0.35cm,0cm);}] coordinates {(V,75.19) (Sh,341.82) (B,683.66) (Su,888.68)};\addlegendentry{Mean AI demand}
\addlegendimage{legend image code/.code={\draw[black,dashed,thick] (0cm,0cm) -- (0.45cm,0cm);}}\addlegendentry{$\varepsilon_i=5\%$}
\node[font=\tiny,text=cBad,anchor=south] at (axis cs:V,100.19) {0.0\%};\node[font=\tiny,text=cBad,anchor=south] at (axis cs:Sh,366.82) {$-$0.05\%};\node[font=\tiny,text=cBad,anchor=south] at (axis cs:B,708.66) {$-$0.2\%};\node[font=\tiny,text=cBad,anchor=south] at (axis cs:Su,913.68) {$-$4.6\%};
\end{axis}
\end{tikzpicture}
\end{minipage}\hfill
\begin{minipage}[b]{0.49\textwidth}
\centering
\begin{tikzpicture}
\begin{axis}[width=\linewidth,
  height=3.7cm,
  ybar=1pt,
  bar width=12pt,
  symbolic x coords={V,Sh,B,Su},
  xtick=data,
  xticklabels={{Valley\\$\times$0.11},{Shoulder\\$\times$0.5},{Busy\\$\times$1.0},{Event\\$\times$1.3}},
  xticklabel style={align=center},
  enlarge x limits=0.15,
  area legend,
  ymajorgrids,
  grid style={black!10},
  ymin=0,
  ymax=58,
  ylabel={Worst-site RAN violation (\%)},
  title={(b) RAN violation probability (label: RAN slices)}]
\addplot[fill=cProp,draw=cProp,point meta=explicit symbolic,nodes near coords,every node near coord/.append style={font=\tiny}] coordinates {(V,1.424)[9] (Sh,2.523)[22] (B,4.056)[33] (Su,4.342)[42]};
\addplot[pattern=north east lines,pattern color=cBase,draw=cBase,point meta=explicit symbolic,nodes near coords,every node near coord/.append style={font=\tiny}] coordinates {(V,23.660)[4] (Sh,41.202)[13] (B,46.701)[23] (Su,41.220)[29]};
\addplot[black,dashed,thick,sharp plot,update limits=false] coordinates {(V,5) (Su,5)};
\end{axis}
\end{tikzpicture}
\end{minipage}

\caption{\CAIR{} and mean-demand reservations at four demand levels of the four-site deployment. (a) Protecting the RAN costs at most 4.6\% of the AI throughput. (b) The mean-demand baseline violates the RAN in 24\% to 47\% of the intervals, while \CAIR{} stays below $\varepsilon_i=5\%$; labels give the reserved RAN slices.}
\label{fig:res-time}
\end{figure*}
\begin{figure*}[t]
\centering
\scalebox{1.2}{\begin{tikzpicture}[font=\scriptsize]
\draw[black!80] (-0.1,0) -- (3.670,0);
\draw[black!80] (-0.1,0) -- (-0.1,4.300);
\draw[black!80] (-0.16,0.000) -- (-0.1,0.000) node[left=0pt,font=\scriptsize] {0};
\draw[black!80] (-0.16,1.400) -- (-0.1,1.400) node[left=0pt,font=\scriptsize] {7};
\draw[black!80] (-0.16,2.800) -- (-0.1,2.800) node[left=0pt,font=\scriptsize] {14};
\draw[black!80] (-0.16,4.200) -- (-0.1,4.200) node[left=0pt,font=\scriptsize] {21};
\node[rotate=90,font=\footnotesize] at (-0.68,2.100) {MIG slices};
\filldraw[fill=cRan!35,draw=cRan,line width=0.4pt] (0.000,0.000) rectangle (0.800,1.000);
\node[text=black,font=\fontsize{6.5}{7}\selectfont,inner sep=0pt] at (0.400,0.500) {RAN 5};
\filldraw[fill=cVis!35,draw=cVis!80!black,line width=0.4pt] (0.000,1.000) rectangle (0.800,1.400);
\node[text=black,font=\fontsize{6.5}{7}\selectfont,inner sep=0pt] at (0.400,1.200) {Vis 2g};
\node[font=\scriptsize,anchor=south,inner sep=1pt] at (0.400,1.400) {4.3\%};
\node[anchor=north,font=\scriptsize,inner sep=2pt] at (0.400,0) {Site 1};
\filldraw[fill=cRan!35,draw=cRan,line width=0.4pt] (0.930,0.000) rectangle (1.730,1.800);
\node[text=black,font=\fontsize{6.5}{7}\selectfont,inner sep=0pt] at (1.330,0.900) {RAN 9};
\filldraw[fill=cVis!35,draw=cVis!80!black,line width=0.4pt] (0.930,1.800) rectangle (1.730,2.600);
\node[text=black,font=\fontsize{6.5}{7}\selectfont,inner sep=0pt] at (1.330,2.200) {Vis 4g};
\filldraw[fill=cAsr!35,draw=cAsr!80!black,line width=0.4pt] (0.930,2.600) rectangle (1.730,2.800);
\node[text=black,font=\fontsize{6.5}{7}\selectfont,inner sep=0pt] at (1.330,2.700) {ASR 1g};
\node[font=\scriptsize,anchor=south,inner sep=1pt] at (1.330,2.800) {4.3\%};
\node[anchor=north,font=\scriptsize,inner sep=2pt] at (1.330,0) {Site 2};
\filldraw[fill=cRan!35,draw=cRan,line width=0.4pt] (1.860,0.000) rectangle (2.660,2.400);
\node[text=black,font=\fontsize{6.5}{7}\selectfont,inner sep=0pt] at (2.260,1.200) {RAN 12};
\filldraw[fill=cAsr!35,draw=cAsr!80!black,line width=0.4pt] (1.860,2.400) rectangle (2.660,2.600);
\node[text=black,font=\fontsize{6.5}{7}\selectfont,inner sep=0pt] at (2.260,2.500) {ASR 1g};
\draw[black!55,dashed] (1.860,2.600) rectangle (2.660,2.800);
\node[font=\scriptsize,anchor=south,inner sep=1pt] at (2.260,2.800) {2.3\%};
\node[anchor=north,font=\scriptsize,inner sep=2pt] at (2.260,0) {Site 3};
\filldraw[fill=cRan!35,draw=cRan,line width=0.4pt] (2.790,0.000) rectangle (3.590,3.200);
\node[text=black,font=\fontsize{6.5}{7}\selectfont,inner sep=0pt] at (3.190,1.600) {RAN 16};
\filldraw[fill=cVis!35,draw=cVis!80!black,line width=0.4pt] (2.790,3.200) rectangle (3.590,4.000);
\node[text=black,font=\fontsize{6.5}{7}\selectfont,inner sep=0pt] at (3.190,3.600) {Vis 4g};
\filldraw[fill=cAsr!35,draw=cAsr!80!black,line width=0.4pt] (2.790,4.000) rectangle (3.590,4.200);
\node[text=black,font=\fontsize{6.5}{7}\selectfont,inner sep=0pt] at (3.190,4.100) {ASR 1g};
\node[font=\scriptsize,anchor=south,inner sep=1pt] at (3.190,4.200) {4.3\%};
\node[anchor=north,font=\scriptsize,inner sep=2pt] at (3.190,0) {Site 4};
\node[anchor=south,font=\footnotesize,align=center] at (1.795,4.500) {(a) C$^2$AIR\\[-1pt]\scriptsize 42 RAN slices, 597 req/s};
\end{tikzpicture}}\hspace{1.2cm}\scalebox{1.2}{\begin{tikzpicture}[font=\scriptsize]
\draw[black!80] (-0.1,0) -- (3.670,0);
\filldraw[fill=cRan!35,draw=cRan,line width=0.4pt] (0.000,0.000) rectangle (0.800,0.800);
\node[text=black,font=\fontsize{6.5}{7}\selectfont,inner sep=0pt] at (0.400,0.400) {RAN 4};
\filldraw[fill=cVis!35,draw=cVis!80!black,line width=0.4pt] (0.000,0.800) rectangle (0.800,1.200);
\node[text=black,font=\fontsize{6.5}{7}\selectfont,inner sep=0pt] at (0.400,1.000) {Vis 2g};
\filldraw[fill=cAsr!35,draw=cAsr!80!black,line width=0.4pt] (0.000,1.200) rectangle (0.800,1.400);
\node[text=black,font=\fontsize{6.5}{7}\selectfont,inner sep=0pt] at (0.400,1.300) {ASR 1g};
\node[text=cBad,font=\scriptsize\bfseries,anchor=south,inner sep=1pt] at (0.400,1.400) {20.1\%};
\node[anchor=north,font=\scriptsize,inner sep=2pt] at (0.400,0) {Site 1};
\filldraw[fill=cRan!35,draw=cRan,line width=0.4pt] (0.930,0.000) rectangle (1.730,1.200);
\node[text=black,font=\fontsize{6.5}{7}\selectfont,inner sep=0pt] at (1.330,0.600) {RAN 6};
\filldraw[fill=cVis!35,draw=cVis!80!black,line width=0.4pt] (0.930,1.200) rectangle (1.730,2.000);
\node[text=black,font=\fontsize{6.5}{7}\selectfont,inner sep=0pt] at (1.330,1.600) {Vis 4g};
\filldraw[fill=cAsr!35,draw=cAsr!80!black,line width=0.4pt] (0.930,2.000) rectangle (1.730,2.200);
\node[text=black,font=\fontsize{6.5}{7}\selectfont,inner sep=0pt] at (1.330,2.100) {ASR 1g};
\filldraw[fill=cLsev!30,draw=cLsev!80!black,line width=0.4pt] (0.930,2.200) rectangle (1.730,2.800);
\node[text=black,font=\fontsize{6.5}{7}\selectfont,inner sep=0pt] at (1.330,2.500) {7B 3g};
\node[text=cBad,font=\scriptsize\bfseries,anchor=south,inner sep=1pt] at (1.330,2.800) {41.2\%};
\node[anchor=north,font=\scriptsize,inner sep=2pt] at (1.330,0) {Site 2};
\filldraw[fill=cRan!35,draw=cRan,line width=0.4pt] (1.860,0.000) rectangle (2.660,1.600);
\node[text=black,font=\fontsize{6.5}{7}\selectfont,inner sep=0pt] at (2.260,0.800) {RAN 8};
\filldraw[fill=cAsr!35,draw=cAsr!80!black,line width=0.4pt] (1.860,1.600) rectangle (2.660,1.800);
\node[text=black,font=\fontsize{6.5}{7}\selectfont,inner sep=0pt] at (2.260,1.700) {ASR 1g};
\filldraw[fill=cLsev!30,draw=cLsev!80!black,line width=0.4pt] (1.860,1.800) rectangle (2.660,2.600);
\node[text=black,font=\fontsize{6.5}{7}\selectfont,inner sep=0pt] at (2.260,2.200) {7B 4g};
\draw[black!55,dashed] (1.860,2.600) rectangle (2.660,2.800);
\node[text=cBad,font=\scriptsize\bfseries,anchor=south,inner sep=1pt] at (2.260,2.800) {41.2\%};
\node[anchor=north,font=\scriptsize,inner sep=2pt] at (2.260,0) {Site 3};
\filldraw[fill=cRan!35,draw=cRan,line width=0.4pt] (2.790,0.000) rectangle (3.590,2.200);
\node[text=black,font=\fontsize{6.5}{7}\selectfont,inner sep=0pt] at (3.190,1.100) {RAN 11};
\filldraw[fill=cVis!35,draw=cVis!80!black,line width=0.4pt] (2.790,2.200) rectangle (3.590,3.000);
\node[text=black,font=\fontsize{6.5}{7}\selectfont,inner sep=0pt] at (3.190,2.600) {Vis 4g};
\filldraw[fill=cAsr!35,draw=cAsr!80!black,line width=0.4pt] (2.790,3.000) rectangle (3.590,3.200);
\node[text=black,font=\fontsize{6.5}{7}\selectfont,inner sep=0pt] at (3.190,3.100) {ASR 1g};
\filldraw[fill=cLsev!30,draw=cLsev!80!black,line width=0.4pt] (2.790,3.200) rectangle (3.590,4.000);
\node[text=black,font=\fontsize{6.5}{7}\selectfont,inner sep=0pt] at (3.190,3.600) {7B 4g};
\draw[black!55,dashed] (2.790,4.000) rectangle (3.590,4.200);
\node[text=cBad,font=\scriptsize\bfseries,anchor=south,inner sep=1pt] at (3.190,4.200) {41.2\%};
\node[anchor=north,font=\scriptsize,inner sep=2pt] at (3.190,0) {Site 4};
\node[anchor=south,font=\footnotesize,align=center] at (1.795,4.500) {(b) Mean-demand baseline\\[-1pt]\scriptsize 29 RAN slices, 626 req/s};
\end{tikzpicture}}

\vspace{2pt}
\scalebox{1.2}{\begin{tikzpicture}[font=\scriptsize]
\filldraw[fill=cRan!35,draw=cRan,line width=0.4pt] (0.00,0) rectangle ++(0.25,0.17); \node[anchor=west] at (0.26,0.085) {RAN};
\filldraw[fill=cVis!35,draw=cVis!80!black,line width=0.4pt] (1.07,0) rectangle ++(0.25,0.17); \node[anchor=west] at (1.33,0.085) {Vision};
\filldraw[fill=cAsr!35,draw=cAsr!80!black,line width=0.4pt] (2.53,0) rectangle ++(0.25,0.17); \node[anchor=west] at (2.79,0.085) {ASR};
\filldraw[fill=cLsev!30,draw=cLsev!80!black,line width=0.4pt] (3.60,0) rectangle ++(0.25,0.17); \node[anchor=west] at (3.86,0.085) {LLM7B};
\draw[black!55,dashed] (4.93,0) rectangle ++(0.25,0.17); \node[anchor=west] at (5.19,0.085) {Unused};
\node[anchor=west,text=cBad,font=\scriptsize\bfseries] at (6.38,0.085) {x\%};\node[anchor=west] at (6.80,0.085) {violation $>\varepsilon_i$};
\end{tikzpicture}}

\caption{RAN and AI placement on the MIG slices of each site of the four-site deployment in the event hour; the label above each bar is the RAN violation probability.}
\label{fig:res-placement}
\end{figure*}
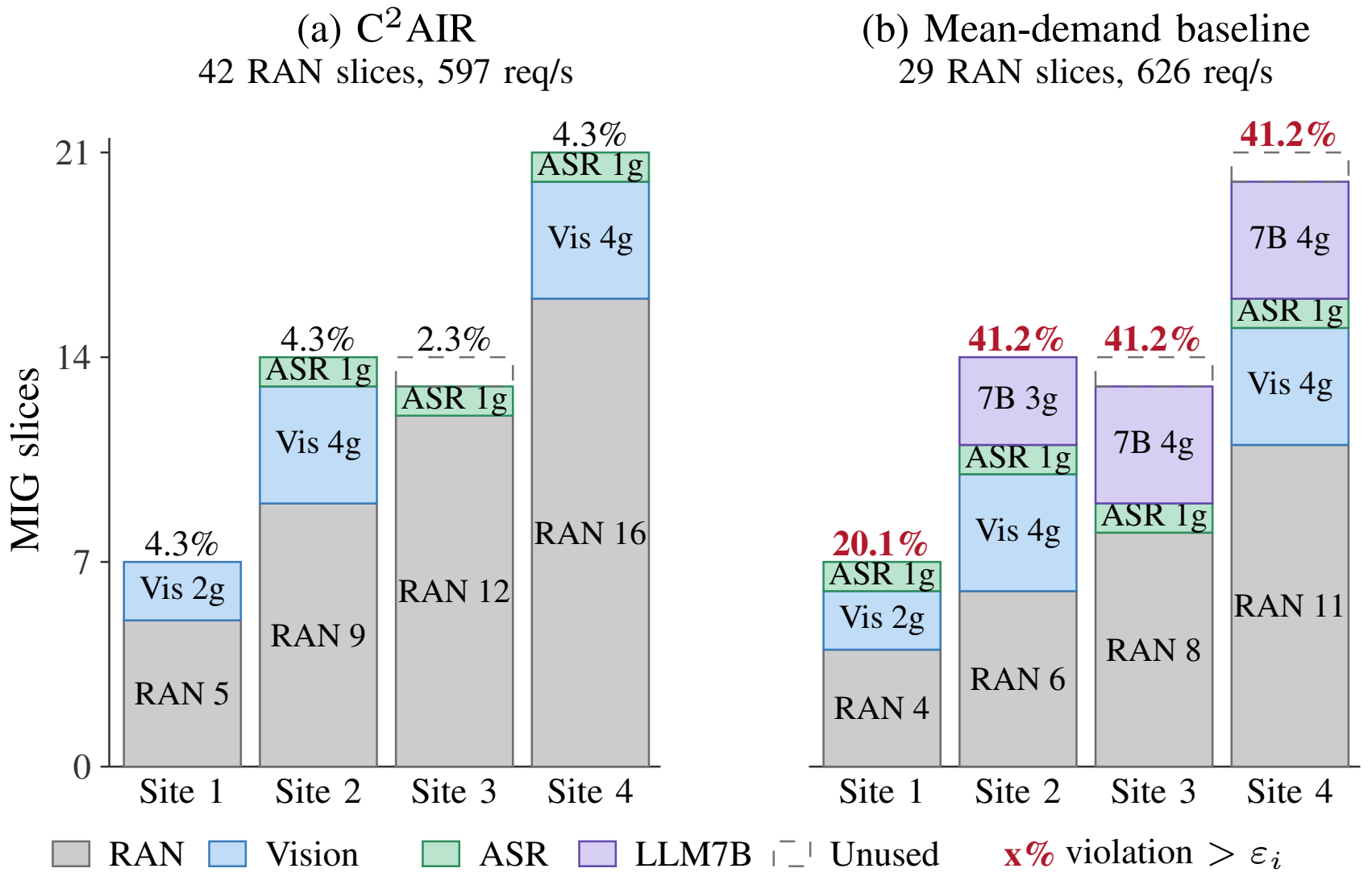
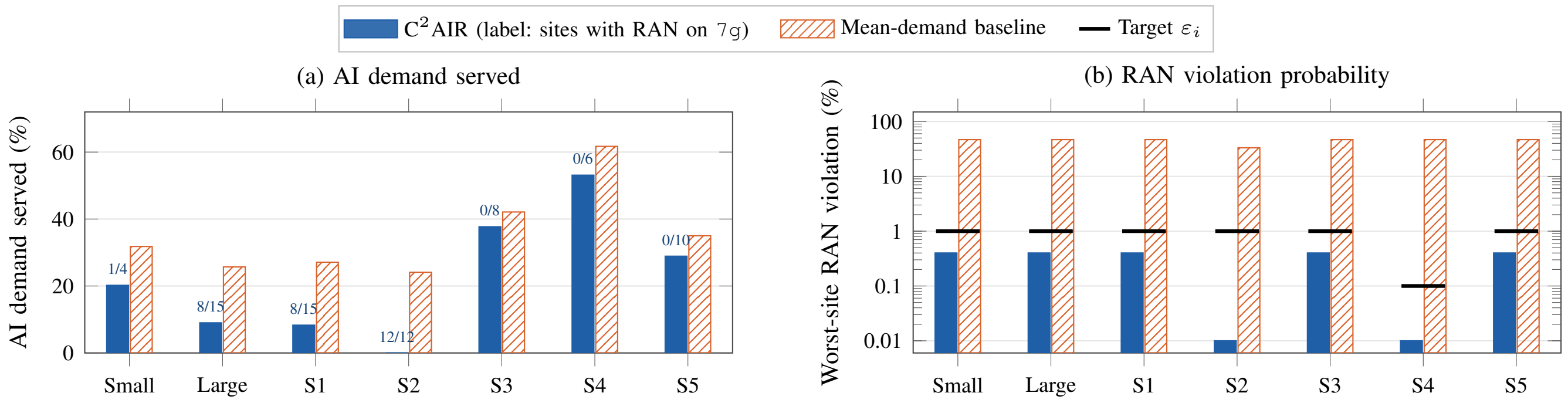
\begin{figure*}[t]
\centering
\ref{lg:scenres}\\[1pt]
\begin{minipage}[b]{0.49\textwidth}
\centering
\begin{tikzpicture}
\begin{axis}[width=\linewidth,
  height=4.3cm,
  ybar=0.6pt,
  bar width=7pt,
  symbolic x coords={Small,Large,S1,S2,S3,S4,S5},
  xtick=data,
  enlarge x limits=0.08,
  area legend,
  ymajorgrids,
  grid style={black!10},
  ymin=0,
  ymax=72,
  ylabel={AI demand served (\%)},
  title={(a) AI demand served},
  legend to name=lg:scenres,
  legend columns=-1,
  legend style={draw=black!25,/tikz/every even column/.append style={column sep=8pt}}]
\addplot[fill=cProp,draw=cProp,point meta=explicit symbolic,nodes near coords,every node near coord/.append style={font=\tiny,text=cProp!70!black}] coordinates {(Small,20.2)[1/4] (Large,9.0)[8/15] (S1,8.3)[8/15] (S2,0.0)[12/12] (S3,37.7)[0/8] (S4,53.1)[0/6] (S5,28.9)[0/10]};
\addlegendentry{C$^2$AIR (label: sites with RAN on \texttt{7g})}
\addplot[pattern=north east lines,pattern color=cBase,draw=cBase] coordinates {(Small,31.8) (Large,25.7) (S1,27.1) (S2,24.1) (S3,42.1) (S4,61.7) (S5,35.0)};\addlegendentry{Mean-demand baseline}
\addlegendimage{legend image code/.code={\draw[black,very thick] (0cm,0cm) -- (0.35cm,0cm);}}\addlegendentry{Target $\varepsilon_i$}
\end{axis}
\end{tikzpicture}
\end{minipage}\hfill
\begin{minipage}[b]{0.49\textwidth}
\centering
\begin{tikzpicture}
\begin{axis}[width=\linewidth,
  height=4.3cm,
  ybar=0.6pt,
  bar width=7pt,
  symbolic x coords={Small,Large,S1,S2,S3,S4,S5},
  xtick=data,
  enlarge x limits=0.08,
  area legend,
  ymajorgrids,
  grid style={black!10},
  ymode=log,
  ymin=0.006,
  ymax=150,
  log origin=infty,
  ylabel={Worst-site RAN violation (\%)},
  title={(b) RAN violation probability},
  ytick={0.01,0.1,1,10,100},
  yticklabels={0.01,0.1,1,10,100}]
\addplot[fill=cProp,draw=cProp] coordinates {(Small,0.4) (Large,0.4) (S1,0.4) (S2,0.01) (S3,0.4) (S4,0.01) (S5,0.4)};
\addplot[pattern=north east lines,pattern color=cBase,draw=cBase] coordinates {(Small,46.7) (Large,46.7) (S1,46.7) (S2,33.19) (S3,46.7) (S4,46.7) (S5,46.7)};
\addplot[only marks,mark=-,mark size=7pt,very thick,black] coordinates {(Small,1.0) (Large,1.0) (S1,1.0) (S2,1.0) (S3,1.0) (S4,0.1) (S5,1.0)};
\end{axis}
\end{tikzpicture}
\end{minipage}
\caption{Larger scenarios at the busy hour. (a) The label above each \CAIR{} bar is the number of sites whose RAN takes the whole GPU (\texttt{7g}). (b) The mean-demand baseline violates the RAN in 33\% to 47\% of the intervals, whereas \CAIR{} stays below the target $\varepsilon_i$ (black marks).}
\label{fig:res-scen-reservation}
\end{figure*}
\begin{figure*}[t]
\centering
\begin{minipage}[b]{0.49\textwidth}
\centering
\begin{tikzpicture}
\begin{axis}[title={(a) RAN threshold and MIG instance},
  width=\linewidth,
  height=4.6cm,
  xmin=4,
  xmax=26,
  ymin=0,
  ymax=142,
  xtick={4,8,12,16,20,24},
  ytick={16,32,48,64,132},
  yticklabels={16 (\texttt{1g}),32 (\texttt{2g}),48 (\texttt{3g}),64 (\texttt{4g}),132 (\texttt{7g})},
  xlabel={Cells per GPU},
  ylabel={SMs},
  legend style={at={(0.02,0.86)},anchor=north west},
  grid=major,
  grid style={black!10}]
\fill[cBad!10] (axis cs:18,0) rectangle (axis cs:26,142);
\node[font=\tiny,text=cBad,align=center,anchor=south] at (axis cs:22.0,5) {RAN takes the\\whole GPU};
\addplot[cProp,thick] coordinates {(4,13.918530405405408) (5,17.457139830508474) (6,21.01982142857143) (7,24.60682167235495) (8,28.218390410958904) (9,31.85478092783505) (10,35.51625000000001) (11,39.20305795847751) (12,42.91546875) (13,46.65375) (14,50.41817307692308) (15,54.20901315789474) (16,58.02654929577465) (17,61.871064487632516) (18,65.74284574468086) (19,69.64218416370107) (20,73.56937500000001) (21,77.52471774193549) (22,81.50851618705036) (23,85.52107851985559) (24,89.56271739130436) (25,93.63375) (26,97.73449817518248)};
\addplot[cProp,thick,dashed] coordinates {(4,12.062726351351353) (5,15.129521186440678) (6,18.217178571428573) (7,21.325912116040957) (8,24.455938356164385) (9,27.60747680412371) (10,30.780750000000005) (11,33.97598356401384) (12,37.19340625) (13,40.43325) (14,43.695750000000004) (15,46.98114473684211) (16,50.28967605633803) (17,53.621589222614844) (18,56.97713297872341) (19,60.35655960854093) (20,63.760125) (21,67.18808870967743) (22,70.64071402877698) (23,74.11826805054152) (24,77.62102173913044) (25,81.14925000000001) (26,84.70323175182482)};
\addplot[black,thick,const plot mark right,no marks] coordinates {(4,16) (5,32) (6,32) (7,32) (8,32) (9,32) (10,48) (11,48) (12,48) (13,48) (14,64) (15,64) (16,64) (17,64) (18,132) (19,132) (20,132) (21,132) (22,132) (23,132) (24,132) (25,132) (26,132)};
\legend{RAN threshold at $\varepsilon_i=0.01$,RAN threshold at $\varepsilon_i=0.05$,RAN MIG instance at $\varepsilon_i=0.01$}
\draw[densely dotted,black!70,thick] (axis cs:20,25) -- (axis cs:20,142);
\node[font=\tiny,anchor=north west,inner sep=1pt] at (axis cs:20.1,120) {S2};
\end{axis}
\end{tikzpicture}
\end{minipage}\hfill
\begin{minipage}[b]{0.49\textwidth}
\centering
\begin{tikzpicture}
\begin{axis}[width=\linewidth,
  height=4.6cm,
  ybar stacked,
  bar width=10pt,
  symbolic x coords={Small,Large,S1,S2,S3,S4,S5},
  xtick=data,
  ymin=0,
  ymax=100,
  ylabel={Sites (\%)},
  enlarge x limits=0.08,
  area legend,
  title={(b) MIG instance of the RAN, busy hour},
  legend style={at={(0.5,-0.2)},anchor=north,legend columns=-1,/tikz/every even column/.append style={column sep=4pt}},
  ymajorgrids,
  grid style={black!10}]
\addplot[fill=cProp!20,draw=black!40] coordinates {(Small,0.0) (Large,0.0) (S1,0.0) (S2,0.0) (S3,0.0) (S4,0.0) (S5,0.0)};\addlegendentry{\texttt{1g}}
\addplot[fill=cProp!45,draw=black!40] coordinates {(Small,25.0) (Large,20.0) (S1,20.0) (S2,0.0) (S3,75.0) (S4,83.33333333333333) (S5,30.0)};\addlegendentry{\texttt{2g}}
\addplot[fill=cProp!70,draw=black!40] coordinates {(Small,25.0) (Large,6.666666666666667) (S1,6.666666666666667) (S2,0.0) (S3,25.0) (S4,16.666666666666668) (S5,60.0)};\addlegendentry{\texttt{3g}}
\addplot[fill=cProp,draw=black!40] coordinates {(Small,25.0) (Large,20.0) (S1,20.0) (S2,0.0) (S3,0.0) (S4,0.0) (S5,10.0)};\addlegendentry{\texttt{4g}}
\addplot[fill=cBad,draw=black!40] coordinates {(Small,25.0) (Large,53.333333333333336) (S1,53.333333333333336) (S2,100.0) (S3,0.0) (S4,0.0) (S5,0.0)};\addlegendentry{\texttt{7g}}
\end{axis}
\end{tikzpicture}
\end{minipage}
\caption{MIG instance granularity on Hopper GPUs at the busy hour. (a) Threshold of the binomial RAN model and the MIG instance the RAN receives; from 18 cells per GPU the threshold at $\varepsilon_i=0.01$ exceeds the \texttt{4g} instance, and the RAN takes the whole GPU (S2 has 20 cells per GPU). (b) Share of the sites of each scenario by the instance its RAN runs on.}
\label{fig:res-cliff}
\end{figure*}
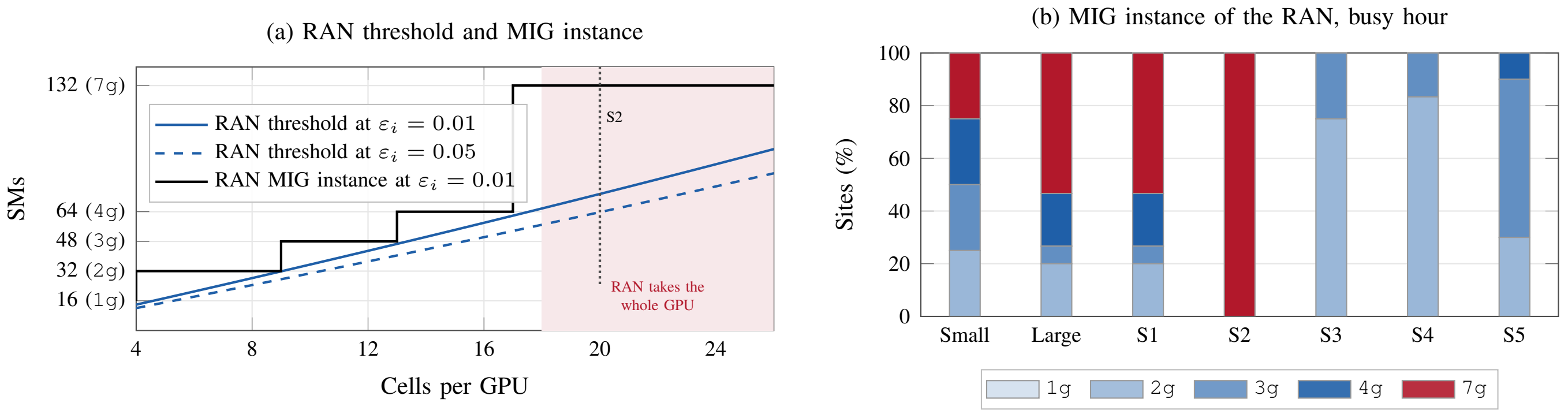
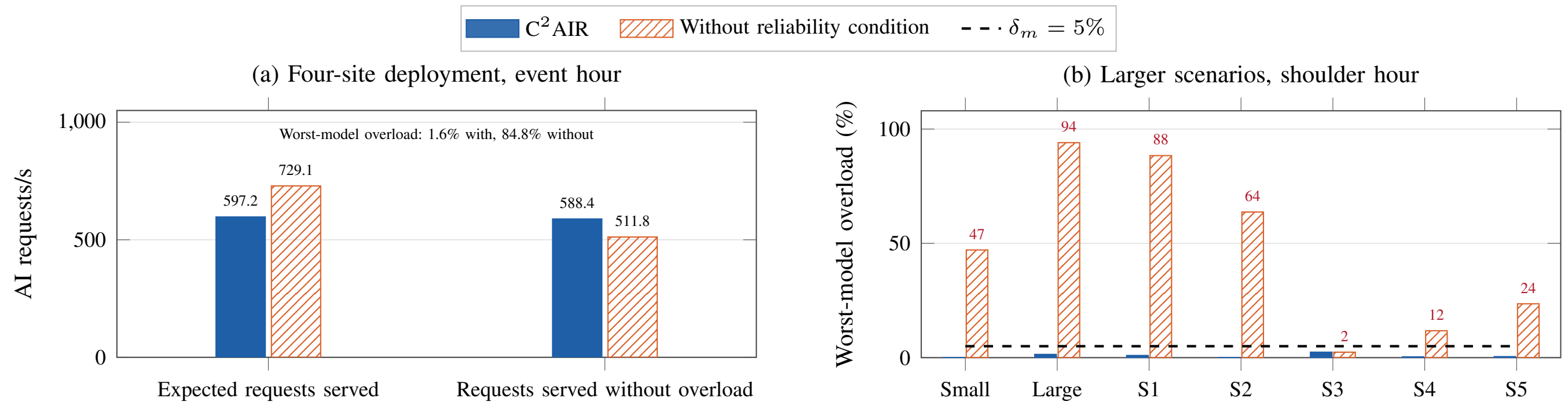
\begin{figure*}[t]
\centering
\ref{lg:elig}\\[1pt]
\begin{minipage}[b]{0.49\textwidth}
\centering
\begin{tikzpicture}
\begin{axis}[width=\linewidth,
  height=4.4cm,
  ybar=2pt,
  bar width=16pt,
  symbolic x coords={E,W},
  xtick=data,
  xticklabels={Expected requests served,Requests served without overload},
  ymin=0,
  ymax=1050,
  ylabel={AI requests/s},
  enlarge x limits=0.45,
  area legend,
  title={(a) Four-site deployment, event hour},
  ymajorgrids,
  grid style={black!10},
  legend to name=lg:elig,
  legend columns=-1,
  legend style={draw=black!25,/tikz/every even column/.append style={column sep=8pt}}]
\addplot[fill=cProp,draw=cProp,nodes near coords,every node near coord/.append style={font=\tiny},nodes near coords={\pgfmathprintnumber[fixed,fixed zerofill,precision=1]{\pgfplotspointmeta}}] coordinates {(E,597.2) (W,588.4)};\addlegendentry{C$^2$AIR}
\addplot[pattern=north east lines,pattern color=cBase,draw=cBase,nodes near coords,every node near coord/.append style={font=\tiny},nodes near coords={\pgfmathprintnumber[fixed,fixed zerofill,precision=1]{\pgfplotspointmeta}}] coordinates {(E,729.1) (W,511.8)};\addlegendentry{Without reliability condition}
\addlegendimage{legend image code/.code={\draw[black,dashed,thick] (0cm,0cm) -- (0.45cm,0cm);}}\addlegendentry{$\delta_m=5\%$}
\node[font=\tiny,align=center,anchor=north] at (rel axis cs:0.5,0.97) {Worst-model overload: 1.6\% with, 84.8\% without};
\end{axis}
\end{tikzpicture}
\end{minipage}\hfill
\begin{minipage}[b]{0.49\textwidth}
\centering
\begin{tikzpicture}
\begin{axis}[width=\linewidth,
  height=4.4cm,
  ybar=0.6pt,
  bar width=7pt,
  symbolic x coords={Small,Large,S1,S2,S3,S4,S5},
  xtick=data,
  ymin=0,
  ymax=108,
  ylabel={Worst-model overload (\%)},
  enlarge x limits=0.08,
  title={(b) Larger scenarios, shoulder hour},
  ymajorgrids,
  grid style={black!10}]
\addplot[fill=cProp,draw=cProp] coordinates {(Small,0.01) (Large,1.41) (S1,0.9400000000000001) (S2,0.04) (S3,2.3800000000000003) (S4,0.33) (S5,0.43)};
\addplot[pattern=north east lines,pattern color=cBase,draw=cBase,nodes near coords,every node near coord/.append style={font=\tiny,text=cBad},nodes near coords={\pgfmathprintnumber[fixed,precision=0]{\pgfplotspointmeta}}] coordinates {(Small,47.099999999999994) (Large,94.05) (S1,88.39) (S2,63.71) (S3,2.3800000000000003) (S4,11.76) (S5,23.56)};
\addplot[black,dashed,thick,sharp plot,update limits=false,legend image code/.code={\draw[black,dashed,thick] (0cm,0cm) -- (0.45cm,0cm);}] coordinates {(Small,5) (S5,5)};
\end{axis}
\end{tikzpicture}
\end{minipage}
\caption{Effect of the AI reliability condition. (a) In the event hour of the four-site deployment, removing the reliability condition raises the expected requests served by 22\% but lowers the requests served without overload by 13\%. (b) In the larger scenarios at the shoulder hour, removing it leaves AI models overloaded in up to 94\% of the intervals, whereas \CAIR{} stays below $\delta_m=5\%$.}
\label{fig:res-eligibility}
\end{figure*}
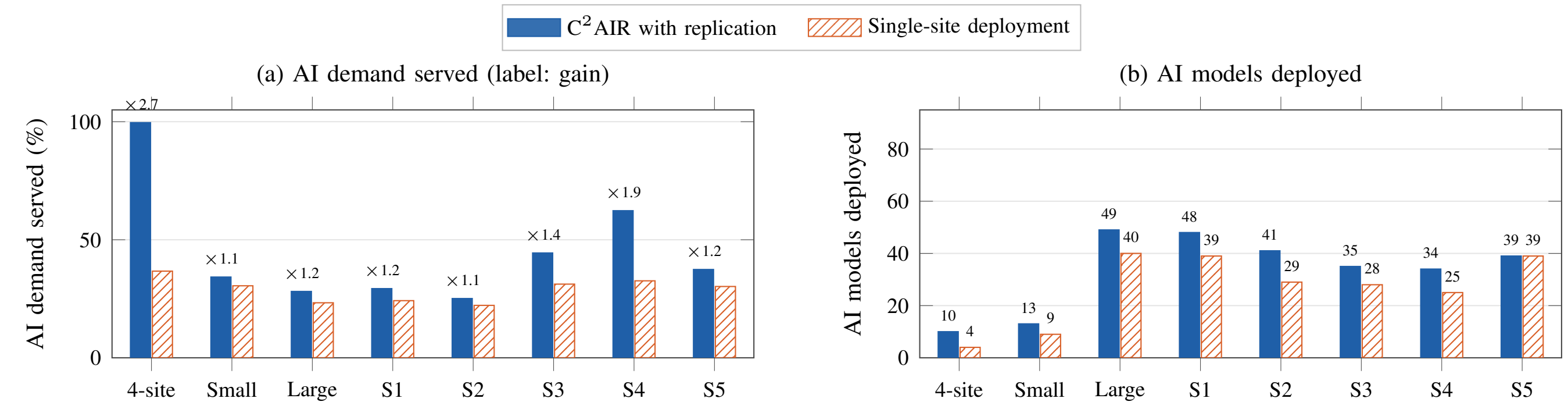
\begin{figure*}[t]
\centering
\ref{lg:repl}\\[1pt]
\begin{minipage}[b]{0.49\textwidth}
\centering
\begin{tikzpicture}
\begin{axis}[width=\linewidth,
  height=4.4cm,
  ybar=0.6pt,
  bar width=6.5pt,
  symbolic x coords={4-site,Small,Large,S1,S2,S3,S4,S5},
  xtick=data,
  ymin=0,
  ymax=105,
  ylabel={AI demand served (\%)},
  enlarge x limits=0.07,
  area legend,
  title={(a) AI demand served (label: gain)},
  ymajorgrids,
  grid style={black!10},
  legend to name=lg:repl,
  legend columns=-1,
  legend style={draw=black!25,/tikz/every even column/.append style={column sep=8pt}}]
\addplot[fill=cProp,draw=cProp,point meta=explicit symbolic,nodes near coords,every node near coord/.append style={font=\tiny,text=black}] coordinates {(4-site,99.6)[$\times$2.7] (Small,34.2)[$\times$1.1] (Large,28.1)[$\times$1.2] (S1,29.3)[$\times$1.2] (S2,25.1)[$\times$1.1] (S3,44.4)[$\times$1.4] (S4,62.3)[$\times$1.9] (S5,37.4)[$\times$1.2]};
\addplot[pattern=north east lines,pattern color=cBase,draw=cBase] coordinates {(4-site,36.7) (Small,30.5) (Large,23.3) (S1,24.2) (S2,22.2) (S3,31.2) (S4,32.6) (S5,30.2)};
\addlegendentry{C$^2$AIR with replication}\addlegendentry{Single-site deployment}
\end{axis}
\end{tikzpicture}
\end{minipage}\hfill
\begin{minipage}[b]{0.49\textwidth}
\centering
\begin{tikzpicture}
\begin{axis}[width=\linewidth,
  height=4.4cm,
  ybar=0.6pt,
  bar width=6.5pt,
  symbolic x coords={4-site,Small,Large,S1,S2,S3,S4,S5},
  xtick=data,
  ymin=0,
  ymax=95,
  ylabel={AI models deployed},
  enlarge x limits=0.07,
  area legend,
  title={(b) AI models deployed},
  nodes near coords,
  every node near coord/.append style={font=\tiny},
  ymajorgrids,
  grid style={black!10}]
\addplot[fill=cProp,draw=cProp] coordinates {(4-site,10.0) (Small,13) (Large,49) (S1,48) (S2,41) (S3,35) (S4,34) (S5,39)};
\addplot[pattern=north east lines,pattern color=cBase,draw=cBase] coordinates {(4-site,4.0) (Small,9) (Large,40) (S1,39) (S2,29) (S3,28) (S4,25) (S5,39)};
\end{axis}
\end{tikzpicture}
\end{minipage}
\caption{Multi-site replication vs. single-site deployment, for the four-site deployment at the busy hour and the larger scenarios at the shoulder hour. (a) AI demand served; labels give the gain of replication. (b) Number of AI models deployed.}
\label{fig:res-replication}
\end{figure*}
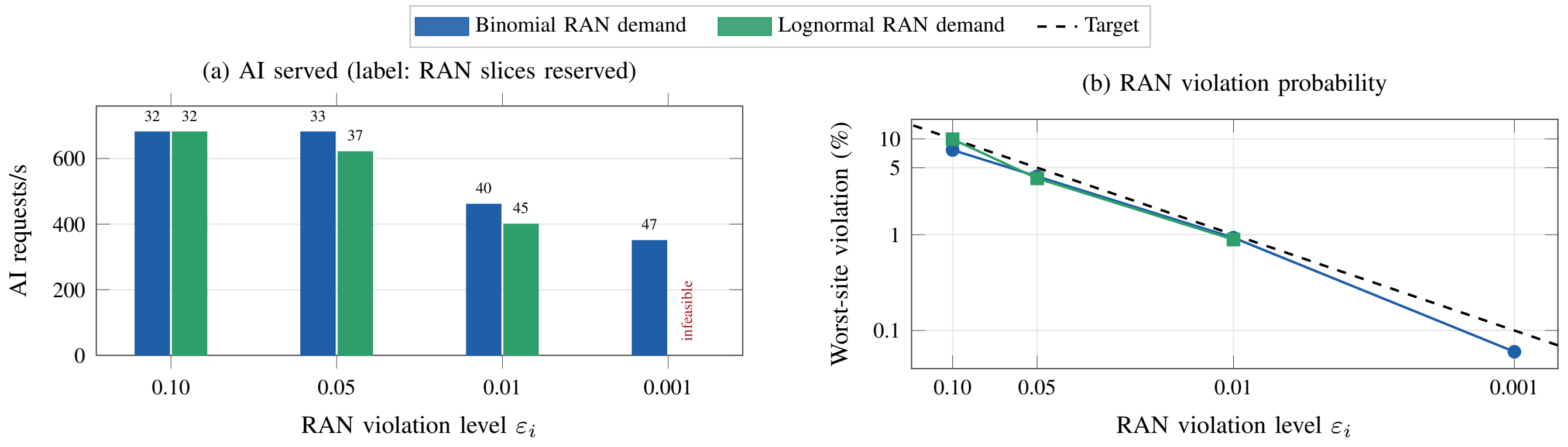
\begin{figure*}[t]
\centering
\ref{lg:eps}\\[1pt]
\begin{minipage}[b]{0.49\textwidth}
\centering
\begin{tikzpicture}
\begin{axis}[width=\linewidth,
  height=4.4cm,
  ybar=1pt,
  bar width=11pt,
  symbolic x coords={0.10,0.05,0.01,0.001},
  xtick=data,
  ymin=0,
  ymax=760,
  ylabel={AI requests/s},
  xlabel={RAN violation level $\varepsilon_i$},
  enlarge x limits=0.15,
  area legend,
  title={(a) AI served (label: RAN slices reserved)},
  ymajorgrids,
  grid style={black!10},
  legend to name=lg:eps,
  legend columns=-1,
  legend style={draw=black!25,/tikz/every even column/.append style={column sep=8pt}}]
\addplot[fill=cProp,draw=cProp,nodes near coords,point meta=explicit symbolic,every node near coord/.append style={font=\tiny}] coordinates {(0.10,680.8)[32] (0.05,680.8)[33] (0.01,460.3)[40] (0.001,349.5)[47]};\addlegendentry{Binomial RAN demand}
\addplot[fill=cBin,draw=cBin,nodes near coords,point meta=explicit symbolic,every node near coord/.append style={font=\tiny}] coordinates {(0.10,680.8)[32] (0.05,620.3)[37] (0.01,399.5)[45]};\addlegendentry{Lognormal RAN demand}
\addlegendimage{legend image code/.code={\draw[black,dashed,thick] (0cm,0cm) -- (0.45cm,0cm);}}\addlegendentry{Target}
\node[font=\tiny,text=cBad,rotate=90,anchor=west] at ([xshift=6pt]axis cs:0.001,10) {infeasible};
\end{axis}
\end{tikzpicture}
\end{minipage}\hfill
\begin{minipage}[b]{0.49\textwidth}
\centering
\begin{tikzpicture}
\begin{axis}[width=\linewidth,
  height=4.4cm,
  xmode=log,
  ymode=log,
  x dir=reverse,
  xmin=0.0007,
  xmax=0.14,
  ymin=0.04,
  ymax=16,
  xlabel={RAN violation level $\varepsilon_i$},
  ylabel={Worst-site violation (\%)},
  title={(b) RAN violation probability},
  xtick={0.001,0.01,0.05,0.1},
  xticklabels={0.001,0.01,0.05,0.10},
  ytick={0.1,1,5,10},
  yticklabels={0.1,1,5,10},
  grid=major,
  grid style={black!10}]
\addplot[black,dashed,thick,domain=0.0007:0.14,samples=2] {100*x};
\addplot[cProp,thick,mark=*,mark size=1.8pt] coordinates {(0.1,7.6499999999999995) (0.05,4.06) (0.01,0.9299999999999999) (0.001,0.06)};
\addplot[cBin,thick,mark=square*,mark size=1.8pt] coordinates {(0.1,9.92) (0.05,3.8699999999999997) (0.01,0.89)};
\end{axis}
\end{tikzpicture}
\end{minipage}
\caption{Effect of the RAN violation level on the four-site deployment for the binomial model of \CAIR{} and the lognormal comparison model. (a) A stricter level reserves more RAN slices and serves less AI; the heavier-tailed lognormal model needs more slices and is infeasible at $\varepsilon_i=0.001$. (b) The violation probability stays at or below the target (dashed) in every feasible case.}
\label{fig:res-eps}
\end{figure*}
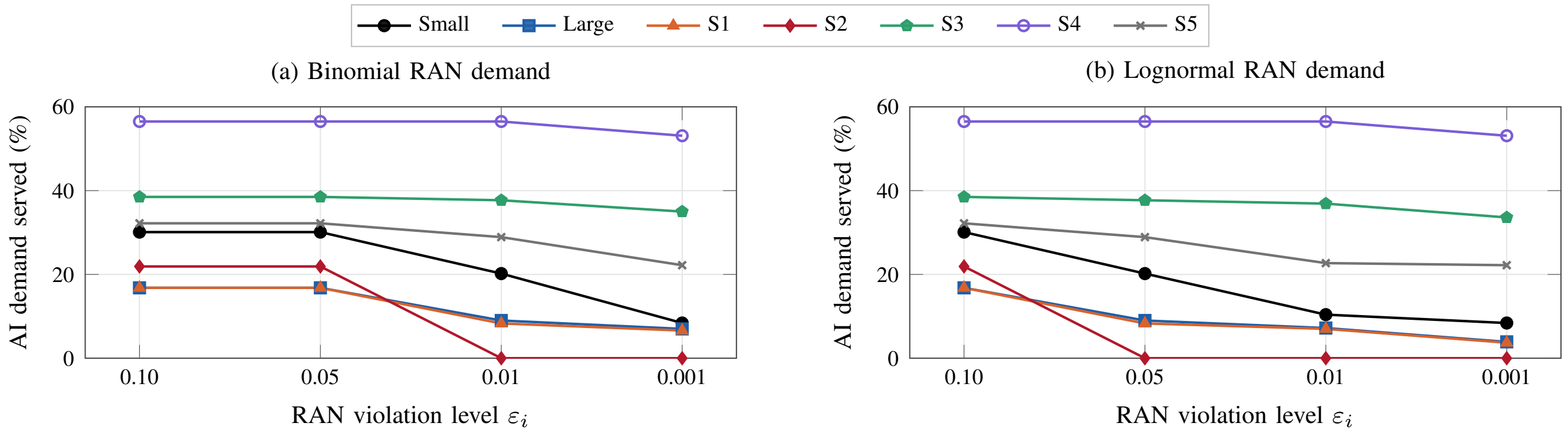
\begin{figure*}[t]
\centering
\ref{lg:epsscen}\\[1pt]
\begin{minipage}[b]{0.49\textwidth}
\centering
\begin{tikzpicture}
\begin{axis}[width=\linewidth,
  height=4.4cm,
  xmin=0.7,
  xmax=4.3,
  xtick={1,2,3,4},
  xticklabels={0.10,0.05,0.01,0.001},
  ymin=0,
  ymax=60,
  xlabel={RAN violation level $\varepsilon_i$},
  ylabel={AI demand served (\%)},
  title={(a) Binomial RAN demand},
  grid=major,
  grid style={black!10},
  legend to name=lg:epsscen,
  legend columns=-1,
  legend style={draw=black!25,/tikz/every even column/.append style={column sep=8pt}}]
\addplot[black,mark=*,thick,mark size=1.7pt,unbounded coords=jump] coordinates {(1,30.1) (2,30.1) (3,20.2) (4,8.4)};\addlegendentry{Small}
\addplot[cProp,mark=square*,thick,mark size=1.7pt,unbounded coords=jump] coordinates {(1,16.8) (2,16.8) (3,9.0) (4,7.0)};\addlegendentry{Large}
\addplot[cBase,mark=triangle*,thick,mark size=1.7pt,unbounded coords=jump] coordinates {(1,16.8) (2,16.8) (3,8.3) (4,6.6)};\addlegendentry{S1}
\addplot[cBad,mark=diamond*,thick,mark size=1.7pt,unbounded coords=jump] coordinates {(1,21.9) (2,21.9) (3,0.0) (4,0.0)};\addlegendentry{S2}
\addplot[cBin,mark=pentagon*,thick,mark size=1.7pt,unbounded coords=jump] coordinates {(1,38.5) (2,38.5) (3,37.7) (4,35.0)};\addlegendentry{S3}
\addplot[cLsev,mark=o,thick,mark size=1.7pt,unbounded coords=jump] coordinates {(1,56.5) (2,56.5) (3,56.5) (4,53.1)};\addlegendentry{S4}
\addplot[black!55,mark=x,thick,mark size=1.7pt,unbounded coords=jump] coordinates {(1,32.2) (2,32.2) (3,28.9) (4,22.2)};\addlegendentry{S5}
\end{axis}
\end{tikzpicture}
\end{minipage}\hfill
\begin{minipage}[b]{0.49\textwidth}
\centering
\begin{tikzpicture}
\begin{axis}[width=\linewidth,
  height=4.4cm,
  xmin=0.7,
  xmax=4.3,
  xtick={1,2,3,4},
  xticklabels={0.10,0.05,0.01,0.001},
  ymin=0,
  ymax=60,
  xlabel={RAN violation level $\varepsilon_i$},
  ylabel={AI demand served (\%)},
  title={(b) Lognormal RAN demand},
  grid=major,
  grid style={black!10}]
\addplot[black,mark=*,thick,mark size=1.7pt,unbounded coords=jump] coordinates {(1,30.1) (2,20.2) (3,10.4) (4,8.4)};
\addplot[cProp,mark=square*,thick,mark size=1.7pt,unbounded coords=jump] coordinates {(1,16.8) (2,9.0) (3,7.2) (4,3.9)};
\addplot[cBase,mark=triangle*,thick,mark size=1.7pt,unbounded coords=jump] coordinates {(1,16.8) (2,8.3) (3,7.0) (4,3.7)};
\addplot[cBad,mark=diamond*,thick,mark size=1.7pt,unbounded coords=jump] coordinates {(1,21.9) (2,0.0) (3,0.0) (4,0.0)};
\addplot[cBin,mark=pentagon*,thick,mark size=1.7pt,unbounded coords=jump] coordinates {(1,38.5) (2,37.7) (3,36.9) (4,33.6)};
\addplot[cLsev,mark=o,thick,mark size=1.7pt,unbounded coords=jump] coordinates {(1,56.5) (2,56.5) (3,56.5) (4,53.1)};
\addplot[black!55,mark=x,thick,mark size=1.7pt,unbounded coords=jump] coordinates {(1,32.2) (2,28.9) (3,22.7) (4,22.2)};
\end{axis}
\end{tikzpicture}
\end{minipage}
\caption{AI demand served in the larger scenarios at the busy hour for different RAN violation levels. Each drop marks sites whose RAN moves to a larger MIG instance; scenarios with few cells per GPU are almost insensitive to $\varepsilon_i$, and the lognormal model drops one step earlier.}
\label{fig:res-eps-scen}
\end{figure*}
\begin{figure*}[t]
\centering
\ref{lg:delta}\\[1pt]
\begin{minipage}[b]{0.49\textwidth}
\centering
\begin{tikzpicture}
\begin{axis}[width=\linewidth,
  height=4.3cm,
  ybar,
  bar width=14pt,
  symbolic x coords={0.20,0.10,0.05,0.01},
  xtick=data,
  xlabel={AI reliability level $\delta_m$},
  enlarge x limits=0.15,
  area legend,
  ymajorgrids,
  grid style={black!10},
  ymin=0,
  ymax=760,
  ylabel={AI requests/s},
  title={(a) AI served},
  legend to name=lg:delta,
  legend columns=-1,
  legend style={draw=black!25,/tikz/every even column/.append style={column sep=8pt}}]
\addplot[fill=cProp,draw=cProp] coordinates {(0.20,680.8) (0.10,680.8) (0.05,680.8) (0.01,482.0)};\addlegendentry{Expected requests served}
\addlegendimage{legend image code/.code={\draw[black,very thick] (0cm,0cm) -- (0.35cm,0cm);}}\addlegendentry{Requests served without overload}
\addlegendimage{legend image code/.code={\draw[cBad,very thick] (0cm,0cm) -- (0.35cm,0cm);}}\addlegendentry{Target $\delta_m$}
\draw[black,very thick] ([xshift=-7pt]axis cs:0.20,676.6) -- ([xshift=7pt]axis cs:0.20,676.6); \draw[black,very thick] ([xshift=-7pt]axis cs:0.10,676.6) -- ([xshift=7pt]axis cs:0.10,676.6); \draw[black,very thick] ([xshift=-7pt]axis cs:0.05,676.6) -- ([xshift=7pt]axis cs:0.05,676.6); \draw[black,very thick] ([xshift=-7pt]axis cs:0.01,481.7) -- ([xshift=7pt]axis cs:0.01,481.7);
\end{axis}
\end{tikzpicture}
\end{minipage}\hfill
\begin{minipage}[b]{0.49\textwidth}
\centering
\begin{tikzpicture}
\begin{axis}[width=\linewidth,
  height=4.3cm,
  xmin=0.7,
  xmax=4.3,
  xtick={1,2,3,4},
  xticklabels={0.20,0.10,0.05,0.01},
  xlabel={AI reliability level $\delta_m$},
  ymin=0,
  ymax=23,
  ylabel={Worst-model overload (\%)},
  title={(b) AI overload probability (red: target $\delta_m$)},
  grid=major,
  grid style={black!10},
  legend style={at={(0.98,0.98)},anchor=north east,legend columns=2,font=\tiny}]
\addplot[black,mark=*,thick,mark size=1.7pt] coordinates {(1,16.400000000000002) (2,8.25) (3,0.06) (4,0.06)};\addlegendentry{Small}
\addplot[cProp,mark=square*,thick,mark size=1.7pt] coordinates {(1,7.61) (2,7.61) (3,1.6) (4,0.01)};\addlegendentry{Large}
\addplot[cBase,mark=triangle*,thick,mark size=1.7pt] coordinates {(1,7.79) (2,7.79) (3,0.43) (4,0.43)};\addlegendentry{S1}
\addplot[cBin,mark=pentagon*,thick,mark size=1.7pt] coordinates {(1,18.8) (2,7.720000000000001) (3,2.77) (4,0.6)};\addlegendentry{S3}
\addplot[cLsev,mark=o,thick,mark size=1.7pt] coordinates {(1,15.98) (2,1.72) (3,1.72) (4,0.08)};\addlegendentry{S4}
\addplot[black!55,mark=x,thick,mark size=1.7pt] coordinates {(1,12.6) (2,2.54) (3,2.54) (4,0.44)};\addlegendentry{S5}
\addplot[cProp,thick,dashed,mark=*,mark options={solid,fill=white},mark size=1.7pt] coordinates {(1,1.0699999999999998) (2,1.0699999999999998) (3,1.0699999999999998) (4,0.29)};\addlegendentry{4-site}
\addplot[cBad,very thick,const plot,no marks,forget plot] coordinates {(0.7,20) (1.5,20) (1.5,10) (2.5,10) (2.5,5) (3.5,5) (3.5,1) (4.3,1)};
\end{axis}
\end{tikzpicture}
\end{minipage}
\caption{Effect of the AI reliability level. (a) On the four-site deployment, only $\delta_m=0.01$ changes the plan; bars give the expected requests served and black marks the requests served without overload. (b) The worst-model overload stays below the target in the four-site deployment and in every larger scenario.}
\label{fig:res-delta}
\end{figure*}
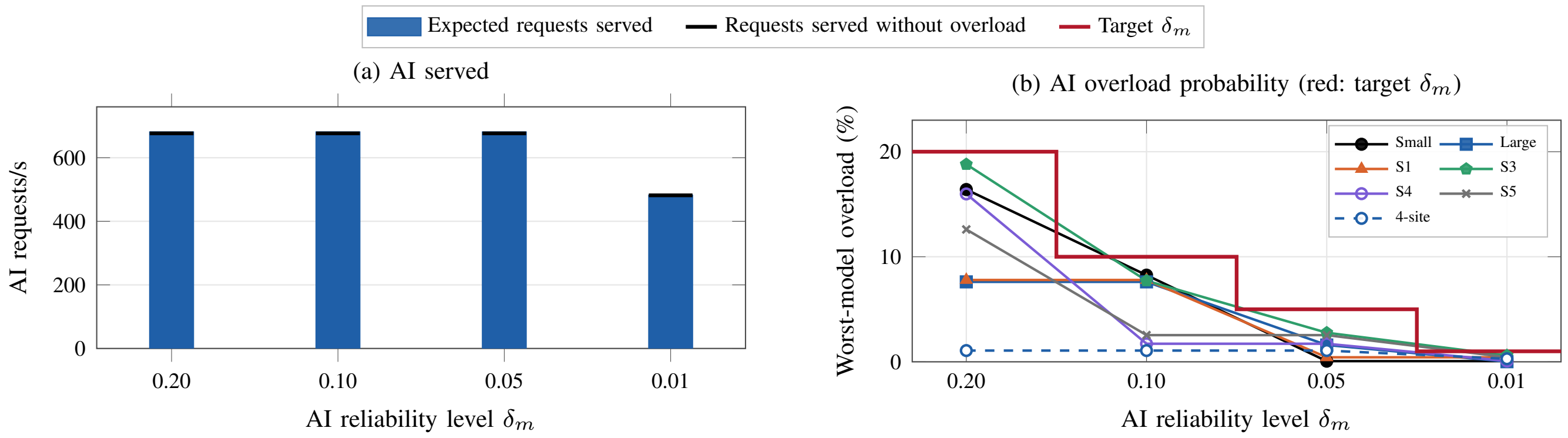
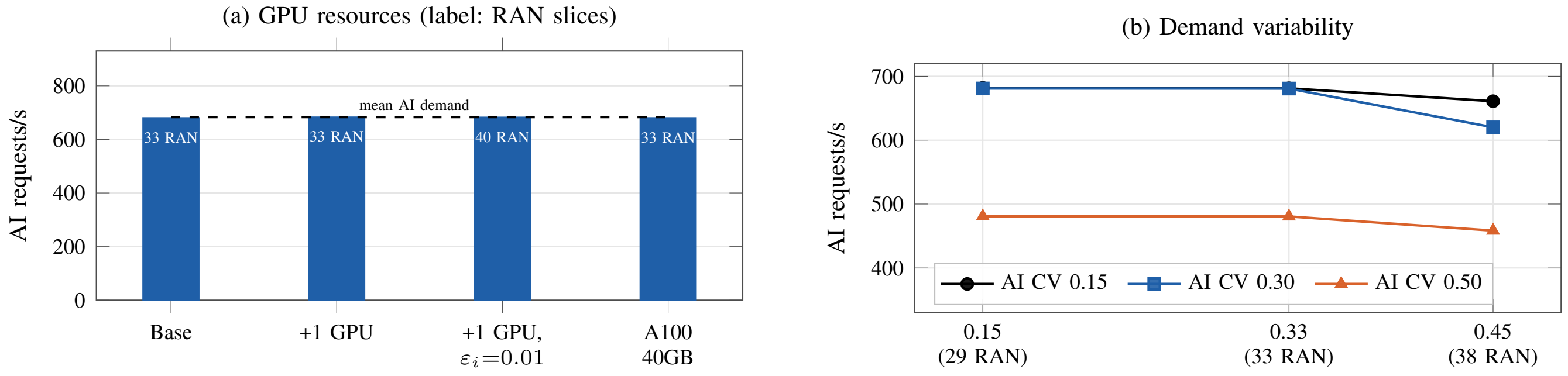
\begin{figure*}[t]
\centering
\begin{minipage}[b]{0.49\textwidth}
\centering
\begin{tikzpicture}
\begin{axis}[width=\linewidth,
  height=4.4cm,
  ybar,
  bar width=18pt,
  symbolic x coords={A,B,C,D},
  xtick=data,
  xticklabels={{Base},{+1 GPU},{+1 GPU,\\$\varepsilon_i{=}0.01$},{A100\\40GB}},
  xticklabel style={align=center},
  ymin=0,
  ymax=930,
  ylabel={AI requests/s},
  enlarge x limits=0.15,
  title={(a) GPU resources (label: RAN slices)},
  nodes near coords,
  point meta=explicit symbolic,
  every node near coord/.append style={font=\tiny,text=white,anchor=north,yshift=-1pt},
  ymajorgrids,
  grid style={black!10}]
\addplot[fill=cProp,draw=cProp] coordinates {(A,680.8)[33 RAN] (B,683.3)[33 RAN] (C,682.8)[40 RAN] (D,680.8)[33 RAN]};
\addplot[black,dashed,thick,sharp plot,update limits=false] coordinates {(A,683.6) (D,683.6)};
\node[font=\tiny,anchor=south,fill=white,inner sep=1pt] at ([xshift=25pt]axis cs:B,695.6) {mean AI demand};
\end{axis}
\end{tikzpicture}
\end{minipage}\hfill
\begin{minipage}[b]{0.49\textwidth}
\centering
\begin{tikzpicture}
\begin{axis}[width=\linewidth,
  height=4.4cm,
  xtick={0.15,0.33,0.45},
  xticklabels={{0.15\\(29 RAN)},{0.33\\(33 RAN)},{0.45\\(38 RAN)}},
  xticklabel style={align=center},
  xmin=0.11,
  xmax=0.49,
  ymin=330,
  ymax=720,
  ylabel={AI requests/s},
  title={(b) Demand variability},
  legend style={at={(0.03,0.03)},anchor=south west,legend columns=-1,/tikz/every even column/.append style={column sep=4pt}},
  grid=major,
  grid style={black!10}]
\addplot[black,mark=*,thick,mark size=1.8pt] coordinates {(0.15,681.8) (0.33,681.2) (0.45,661.0)};
\addplot[cProp,mark=square*,thick,mark size=1.8pt] coordinates {(0.15,680.9) (0.33,680.8) (0.45,620.1)};
\addplot[cBase,mark=triangle*,thick,mark size=1.8pt] coordinates {(0.15,480.7) (0.33,480.6) (0.45,458.4)};
\legend{AI CV 0.15,AI CV 0.30,AI CV 0.50}
\node[font=\scriptsize,anchor=north] at (axis cs:0.3,330) {};
\end{axis}
\end{tikzpicture}
\end{minipage}
\caption{GPU resources and demand variability on the four-site deployment. (a) One extra GPU per site serves 99.9\% of the AI demand even at $\varepsilon_i=0.01$, whereas halving the GPU memory has no effect; labels give the RAN slices and the dashed line marks the mean AI demand. (b) A burstier AI demand costs more throughput than a burstier RAN demand.}
\label{fig:res-resources}
\end{figure*}
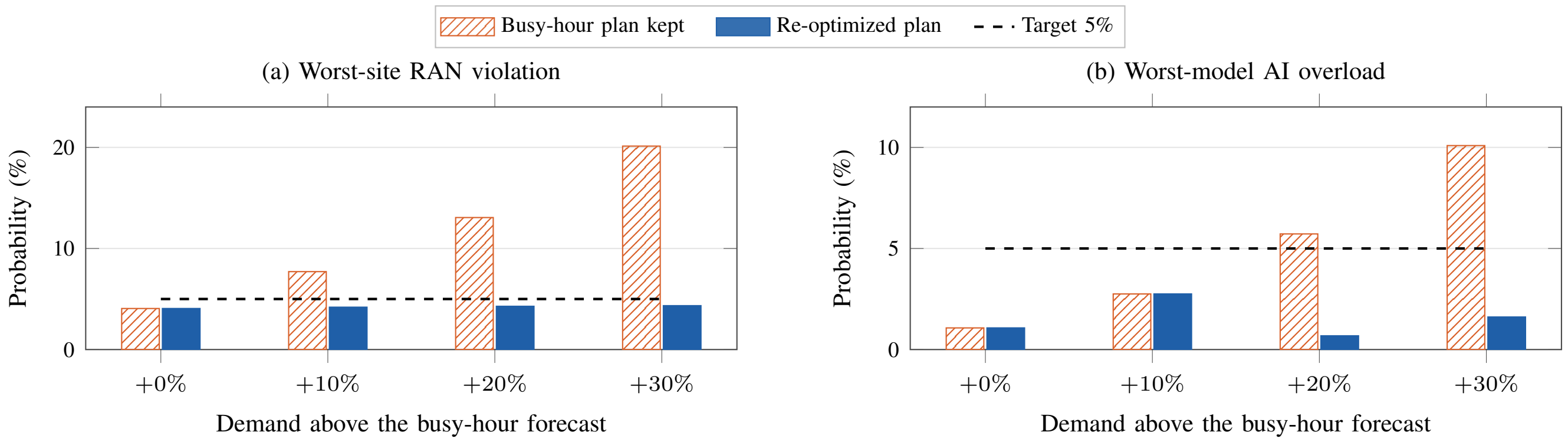
\begin{figure*}[t]
\centering
\ref{lg:growth}\\[1pt]
\begin{minipage}[b]{0.49\textwidth}
\centering
\begin{tikzpicture}
\begin{axis}[width=\linewidth,
  height=4.3cm,
  ybar=1pt,
  bar width=12pt,
  symbolic x coords={G0,G1,G2,G3},
  xtick=data,
  xticklabels={$+0\%$,$+10\%$,$+20\%$,$+30\%$},
  ymin=0,
  ymax=24,
  ylabel={Probability (\%)},
  xlabel={Demand above the busy-hour forecast},
  enlarge x limits=0.15,
  title={(a) Worst-site RAN violation},
  area legend,
  ymajorgrids,
  grid style={black!10},
  legend to name=lg:growth,
  legend columns=-1,
  legend style={draw=black!25,/tikz/every even column/.append style={column sep=8pt}}]
\addplot[pattern=north east lines,pattern color=cBase,draw=cBase] coordinates {(G0,4.06) (G1,7.71) (G2,13.059999999999999) (G3,20.13)};\addlegendentry{Busy-hour plan kept}
\addplot[fill=cProp,draw=cProp] coordinates {(G0,4.06) (G1,4.19) (G2,4.279999999999999) (G3,4.34)};\addlegendentry{Re-optimized plan}
\addplot[black,dashed,thick,sharp plot,update limits=false,legend image code/.code={\draw[black,dashed,thick] (0cm,0cm) -- (0.45cm,0cm);}] coordinates {(G0,5) (G3,5)};\addlegendentry{Target 5\%}
\end{axis}
\end{tikzpicture}
\end{minipage}\hfill
\begin{minipage}[b]{0.49\textwidth}
\centering
\begin{tikzpicture}
\begin{axis}[width=\linewidth,
  height=4.3cm,
  ybar=1pt,
  bar width=12pt,
  symbolic x coords={G0,G1,G2,G3},
  xtick=data,
  xticklabels={$+0\%$,$+10\%$,$+20\%$,$+30\%$},
  ymin=0,
  ymax=12,
  ylabel={Probability (\%)},
  xlabel={Demand above the busy-hour forecast},
  enlarge x limits=0.15,
  title={(b) Worst-model AI overload},
  area legend,
  ymajorgrids,
  grid style={black!10}]
\addplot[pattern=north east lines,pattern color=cBase,draw=cBase] coordinates {(G0,1.0699999999999998) (G1,2.75) (G2,5.72) (G3,10.09)};
\addplot[fill=cProp,draw=cProp] coordinates {(G0,1.0699999999999998) (G1,2.75) (G2,0.6799999999999999) (G3,1.6099999999999999)};
\addplot[black,dashed,thick,sharp plot,update limits=false,legend image code/.code={\draw[black,dashed,thick] (0cm,0cm) -- (0.45cm,0cm);}] coordinates {(G0,5) (G3,5)};
\end{axis}
\end{tikzpicture}
\end{minipage}
\caption{Busy-hour plan kept under unplanned demand growth vs. a re-optimized plan on the four-site deployment. A plan made for the forecast loses its RAN and AI guarantees as soon as the demand grows, whereas re-optimization restores them.}
\label{fig:res-growth}
\end{figure*}

\subsection{Larger Deployment Scenarios}
\label{subsec:setup-scenarios}

The larger scenarios use one Hopper-class GPU per site: an H200 (141~GB), a GH200 (96~GB), or an H100 (80~GB). Each GPU has seven slices of 16 SMs, and its \texttt{7g} instance has all 132 SMs~\cite{NVIDIA_MIG}. Table~\ref{tab:eval-scenarios} summarizes the seven scenarios. Small and Large mix urban and residential sites. S1 Urban is a dense downtown. S2 Operator is an operator RAN pool with 20 cells of 100~MHz per GH200, as in a commercial outdoor AI-RAN trial~\cite{SoftBank2024GH200}. S3 Campus is a university campus dominated by chat, coding, transcription, and embedding services. S4 Factory is a private 5G network for camera-based inspection with the stricter target $\varepsilon_i=0.001$. S5 Residential is a suburb with consumer chat and image generation.

The services are drawn from small, mid-size, and large LLMs, code and vision-language models, image generation, speech, vision, and NLP encoders. Each service has a per-slice rate $\mu_1$ and a scaling exponent $\beta$, so that $\mu_{m,k}=\mu_1 s_k^{\beta}$. The AI demand of a site is split among its services by a Zipf popularity with exponent 0.8, weighted by service class in S3 to S5. The mean RAN demand in SMs follows the Beaver timing model~\cite{Yao2026Beaver} at a busy-hour utilization of 0.5, which gives 15.8~SMs for 8 cells, 32.5~SMs for 16 cells, and 41.2~SMs for 20 cells. Around this mean, the RAN demand is binomial with $\mathrm{CV}_{\mathrm{RAN}}=0.33$ at the busy hour, and the AI demand is Gamma with $\mathrm{CV}_{\mathrm{AI}}=0.30$. Demands are counted in requests per hour.

Two hardware rules of the Hopper GPUs complete \CAIR{} for these scenarios. First, the RAN runs on a single MIG instance of 1, 2, 3, 4, or 7 slices, so $\phi_i$ takes only the values 16, 32, 48, 64, and 132~SMs, and the memory of that instance counts in the memory budget; the level encoding of Section~\ref{subsec:capacity-mapping} leaves out the other levels. Second, each profile respects its maximum number of instances per GPU.

\begin{table}[t]
\caption{Larger Deployment Scenarios. Binaries Is the Number of Binary Variables at the Busy Hour.}
\label{tab:eval-scenarios}
\centering
\footnotesize
\setlength{\tabcolsep}{3.2pt}
\resizebox{\columnwidth}{!}{%
\begin{tabular}{llccccr}
\toprule
Scenario & GPU & $\varepsilon_i$ & Sites & Cells & Cells/GPU & AI services / Binaries\\
\midrule
Small & H200 & 0.01 & 4 & 56 & 8--20 & 10 / 190\\
Large & H200 & 0.01 & 15 & 250 & 6--24 & 40 / 3{,}297\\
S1 Urban & H200 & 0.01 & 15 & 250 & 6--24 & 40 / 3{,}290\\
S2 Operator & GH200 & 0.01 & 12 & 240 & 20 & 40 / 2{,}470\\
S3 Campus & H100 & 0.01 & 8 & 64 & 6--12 & 40 / 1{,}561\\
S4 Factory & H100 & 0.001 & 6 & 34 & 4--8 & 26 / 879\\
S5 Residential & H200 & 0.01 & 10 & 100 & 6--16 & 40 / 2{,}230\\
\bottomrule
\end{tabular}}
\end{table}

\subsection{Baselines, Metrics, and Solver}
\label{subsec:setup-metrics}

\textit{Baselines:} each baseline removes one feature of \CAIR{} and keeps everything else.
\begin{itemize}\setlength{\itemsep}{0pt}
  \item \textbf{Mean-demand reservation} replaces the chance constraint with $\phi_i(y_i)\geq\E[\widetilde d_{i,t}]$ for all $i\in\Iset$, so it ignores the variability of the RAN demand.
  \item \textbf{No reliability condition} restricts the profiles only by memory, so an AI model may run on a profile that is too small for its demand.
  \item \textbf{Single-site deployment} places each service at one site at most, $\sum_{i\in\Iset}\sum_{k\in\Kset^{\mathrm{AI}}_{m,i,t}}x_{m,i,k}\leq1$ for all $m\in\Mset$, so it shows the value of replication.
\end{itemize}

\textit{Metrics:}
\begin{itemize}\setlength{\itemsep}{0pt}
  \item \textbf{Expected requests served:} \eqref{eq:objective} in \mbox{requests/s}, also given as a share of the mean AI demand.
  \item \textbf{Requests served without overload:} the throughput delivered within the latency bound, which counts only the intervals in which an AI model keeps up with its demand. It is computed as $\sum_{i\in\Iset}\sum_{m\in\Mset}\sum_{k\in\Kset^{\mathrm{AI}}_{m,i,t}}u_{m,i,k,t}\,x_{m,i,k}$, where $u_{m,i,k,t}=\kappa_{m,i}\theta_{m,i}F_{\Gamma(\kappa_{m,i}+1,\theta_{m,i})}(\mu_{m,k})$ is the first term of \eqref{eq:ai-served}.
  \item \textbf{Worst-site RAN violation probability:} the probability $\Prb\big(\widetilde d_{i^\star,t}>\phi_{i^\star}(y_{i^\star})\big)$ at the site $i^\star=\arg\max_{i\in\Iset}\Prb\big(\widetilde d_{i,t}>\phi_i(y_i)\big)$. It shows whether the RAN target $\varepsilon_i$ is met.
  \item \textbf{Worst-model overload probability:} $\max\{\Prb(\widetilde\lambda_{m,i,t}>\mu_{m,k}):x_{m,i,k}=1\}$. It shows whether the AI target $\delta_m$ is met.
\end{itemize}

\textit{Solver:} \CAIR{} is implemented with the docplex API and solved with IBM ILOG CPLEX~22.2~\cite{IBM_CPLEX}. Scenario instances above 1{,}000 variables, which exceed the CPLEX Community Edition limit, are solved with HiGHS~\cite{Huangfu2018HiGHS} to a zero optimality gap. A negligible penalty on the total reservation, $10^{-4}$ per slice, breaks ties in favor of the smallest RAN reservation.
%
\section{Experimental Results}
\label{sec:results}

Each experiment changes one setting of \CAIR{} or of the deployment while keeping everything else at the default of Section~\ref{sec:evaluation}, and each one tests one design choice of \CAIR{}. Table~\ref{tab:experiments} lists what every subsection changes, what it is compared with, and which design choice it tests. Within a subsection, we report the four-site deployment first, where every decision can be inspected, and then the larger scenarios, which test whether the observation carries over to a larger scale and to Hopper GPUs. Every subsection ends with the message it brings to the evaluation of \CAIR{}. All probabilities are computed exactly from the demand distributions.

\begin{table}[t]
\caption{Experiments and the Design Choice of \CAIR{} That Each One Tests}
\label{tab:experiments}
\centering
\footnotesize
\setlength{\tabcolsep}{2.5pt}
\renewcommand{\arraystretch}{1.1}
\begin{tabular}{>{\raggedright\arraybackslash}p{0.08\columnwidth}>{\raggedright\arraybackslash}p{0.28\columnwidth}>{\raggedright\arraybackslash}p{0.24\columnwidth}>{\raggedright\arraybackslash}p{0.30\columnwidth}}
\toprule
Sec. & What changes & Compared with & Design choice tested\\
\midrule
\ref{subsec:res-reservation} & RAN reservation rule, at four demand levels & Mean-demand reservation & RAN chance constraint; MIG instance levels\\
\ref{subsec:res-eligibility} & Profiles an AI model may use & No reliability condition & AI reliability condition\\
\ref{subsec:res-replication} & Number of sites per service & Single-site deployment & Multi-site replication\\
\ref{subsec:res-levels} & $\varepsilon_i$ and the RAN demand distribution & Lognormal RAN demand & Chance constraint and its deterministic equivalent\\
\ref{subsec:res-delta} & $\delta_m$ & Looser and stricter $\delta_m$ & AI reliability condition\\
\ref{subsec:res-resources} & GPUs, GPU memory, demand CVs & Base deployment & Use of \CAIR{} for capacity planning\\
\ref{subsec:res-classes} & Demand level and service class (scenarios) & Mean-demand reservation & Re-optimization in every interval\\
\ref{subsec:res-growth} & Demand above the forecast & Busy-hour plan kept & Limit of single-stage planning\\
\ref{subsec:res-time} & Instance size & None & MILP tractability\\
\bottomrule
\end{tabular}
\end{table}

\subsection{Mean-Demand vs. Chance-Constrained RAN Reservation}
\label{subsec:res-reservation}

The first question is whether the RAN needs a probabilistic reservation at all, or whether reserving for its mean demand is enough. The mean-demand baseline keeps the AI side of \CAIR{} unchanged and replaces only the right-hand side of the RAN constraint, so any difference comes from the reservation rule. Fig.~\ref{fig:res-time} re-optimizes both reservations of the four-site deployment at the four demand levels of the day. \CAIR{} reserves 9, 22, 33, and 42 of the 56 slices and keeps the worst-site violation between 1.4\% and 4.3\%, below $\varepsilon_i=5\%$. At the busy hour, for example, the right-hand sides of \eqref{eq:dep-binomial} are 55.7, 97.4, 125.2, and 180.9~SMs, about 1.55 times the mean demand, so the sites reserve 4, 7, 9, and 13 slices, and the remaining 23 slices serve 680.8 requests/s, 99.6\% of the mean AI demand. The mean-demand baseline reserves only 4, 13, 23, and 29 slices and violates the RAN in 24\% to 47\% of the intervals at every level, roughly one interval out of three. The price of protection grows with the load: it is below 0.1\% at the valley and shoulder hours, 0.2\% at the busy hour, and 4.6\% in the event hour. At low load the slices that the RAN tail needs are spare anyway; at high load they must be taken from AI.

Fig.~\ref{fig:res-placement} shows where this price comes from in the event hour. \CAIR{} reserves 42 of the 56 slices and admits only Vision and ASR models, whereas the baseline reserves 29 slices and also hosts LLM7B at three sites at the cost of 20\% to 41\% RAN violation. The baseline's extra 4.8\% of AI throughput (625.8 against 597.2 requests/s) is therefore bought with RAN slots that miss their deadline in up to two intervals out of five.

The larger scenarios lead to the same conclusion (Fig.~\ref{fig:res-scen-reservation}). At the busy hour, the mean-demand plans serve more AI in every scenario, but their worst site violates the RAN in 33\% to 47\% of the intervals. \CAIR{} keeps every site at or below 0.4\%, inside the target of 1\% (0.1\% in S4). Unlike the four-site deployment, the AI cost of protection is large in the RAN-heavy scenarios: Large and S1 serve 9.0\% and 8.3\% of the AI demand instead of 25.7\% and 27.1\%, and S2 serves none instead of 24.1\%.

The reason is the granularity of MIG instances, illustrated in Fig.~\ref{fig:res-cliff}. On a Hopper GPU the RAN runs on one instance, and there is no instance between four slices (64~SMs) and the whole GPU (132~SMs). At $\varepsilon_i=0.01$ the binomial threshold is 1.78 times the mean demand and exceeds 64~SMs from 18 cells per GPU, so the RAN then takes the whole GPU and leaves nothing for AI. This happens at 1 of the 4 sites of Small, 8 of the 15 sites of Large and S1, and all 12 sites of S2, which has 20 cells per GPU (Fig.~\ref{fig:res-cliff}(b)). Sites with fewer cells keep most of their GPU for AI, so S3 Campus serves 37.7\% and S4 Factory 53.1\% of the AI demand. This hardware rule entered \CAIR{} only through the admissible levels of the encoding in Section~\ref{subsec:capacity-mapping}, without any change to the formulation. A finer instance between \texttt{4g} and \texttt{7g}, or a RAN that can span two instances, would recover most of the AI capacity lost in S2.

\emph{Takeaway:} reserving for the mean RAN demand fails the RAN in roughly one interval out of three, in every deployment and at every load. The chance constraint of \CAIR{} meets the target everywhere, at a cost to AI that is small when slices are spare and that is set by the demand tail and the MIG instance sizes when they are not, both of which \CAIR{} accounts for exactly.

\subsection{Placement without vs. with the AI Reliability Condition}
\label{subsec:res-eligibility}

The reliability condition decides which profiles a service may use. The baseline removes it and keeps only the memory check, so the optimizer may place an AI model on a profile that is too small for its local demand; the RAN side is unchanged. Because the objective counts the capacity of an overloaded model as served, this baseline always expects at least as many requests as \CAIR{}, and the question is whether these requests are actually delivered. We therefore compare the expected requests served with the requests served without overload.

At the busy hour of the four-site deployment both plans coincide, because enough slices remain to give every model a reliable profile. In the event hour, when slices are scarce (Fig.~\ref{fig:res-eligibility}(a)), the plan without the reliability condition runs Vision on \texttt{1g} at Site~1 and on \texttt{2g} at Site~3. It expects 729.1 requests/s, 22\% more than \CAIR{}, but Vision at Site~3 is overloaded in 85\% of the intervals, against 1.6\% for the worst model of \CAIR{}. Counting only intervals without overload, it serves 13\% fewer requests (511.8 against 588.4 requests/s), so the extra expected throughput is never delivered within the latency bound.

The larger scenarios confirm that the extra throughput comes from overloaded AI models (Fig.~\ref{fig:res-eligibility}(b)). At the shoulder hour, the worst AI model without the reliability condition is overloaded in 64\% to 94\% of the intervals in Large, S1, and S2, and at the busy hour in up to 98\%. Counted without overload, these plans serve less than \CAIR{} in Large, S1, S2, and S4. In Small and S5 they gain about one percentage point, but only by running AI models that are overloaded in 47\% and 24\% of the intervals.

\emph{Takeaway:} without the reliability condition, the expected throughput overstates what the AI models deliver, because the extra requests come from models that are overloaded most of the time. The reliability condition of \CAIR{} costs nothing when slices are available and, when they are scarce, turns the expected throughput into throughput delivered within the latency bound.

\subsection{Single-Site Deployment vs. Multi-Site Replication}
\label{subsec:res-replication}

Because an AI model serves only the demand of its own site, a service deployed at one site leaves the demand of all other sites unserved. The single-site baseline adds one constraint per service that allows at most one AI model across all sites, and everything else, including the RAN reservation, is unchanged. Fig.~\ref{fig:res-replication} compares the two policies. In the four-site deployment, single-site placement runs only four AI models, Vision at Site~3 and the other three services at Site~4. It leaves 11 slices unused and serves 250.7 instead of 680.8 requests/s, so replication serves 2.7 times more with the same RAN protection.

In the larger scenarios at the shoulder hour, the gain ranges from 1.1 times in Small and S2 to 1.9 times in S4 Factory. Fig.~\ref{fig:res-replication}(b) explains the range. Single-site deployment admits more distinct services, up to 40 in Large, but each of them serves one site only, so it runs fewer AI models than \CAIR{} (40 against 49 in Large) or the same number with less demand behind each one (39 in S5). The gain is largest where the same services are requested at many sites, as the inspection cameras of S4 and the campus services of S3, and smallest where each site already fills its GPU with its own popular services, so that forbidding replication mostly changes which services run.

\emph{Takeaway:} because AI demand is local, the value of a placement lies in matching services to the sites where they are requested. Replication, which \CAIR{} allows with one profile per site, serves up to 2.7 times more AI requests without weakening the RAN protection.

\subsection{RAN Violation Level and Demand Tail}
\label{subsec:res-levels}

The violation level $\varepsilon_i$ is the main lever of an operator between RAN safety and AI throughput, and it enters \CAIR{} only through the right-hand side of \eqref{eq:dep-binomial}. Fig.~\ref{fig:res-eps}(a) varies it on the four-site deployment. With the binomial model, the RAN takes 32, 33, 40, and 47 slices for $\varepsilon_i=0.10$, 0.05, 0.01, and 0.001, and AI served is 680.8 requests/s at the two looser levels, 460.3 requests/s at 0.01, and 349.5 requests/s at 0.001. Moving from 5\% to 1\% therefore costs 32\% of the AI throughput, because Vision at Site~3 loses its slices. Fig.~\ref{fig:res-eps}(b) shows that every plan keeps its violation probability at or below the target, and the realized violation lies below the target only because the reservation comes in whole slices; the deterministic equivalent itself adds no safety margin.

The second question is how much the answer depends on the distribution assumed for the RAN demand. The lognormal model of Section~\ref{subsec:lognormal} is not a baseline but a second demand model with the same mean and CV and a heavier upper tail. It needs more slices at the same level (37 instead of 33 at $\varepsilon_i=0.05$, 45 instead of 40 at 0.01) and serves 620.3 and 399.5 requests/s. At $\varepsilon_i=0.001$ it is infeasible, because the 99.9th percentile of its demand is 208~SMs at Site~3 and 300~SMs at Site~4, more than the 196 and 294~SMs of their GPUs.

In the larger scenarios (Fig.~\ref{fig:res-eps-scen}), relaxing $\varepsilon_i$ helps most where the RAN sits near the boundary between the \texttt{4g} and \texttt{7g} instances. At $\varepsilon_i=0.05$ every S2 RAN fits on \texttt{4g}, and 21.9\% of the AI demand is served instead of none; Small and Large reach 30.1\% and 16.8\%. Scenarios with few cells per GPU change little over the whole range: S3 stays between 35.0\% and 38.5\%, and S4 between 53.1\% and 56.5\%. The lognormal model moves more RANs to \texttt{7g} at the same level, so its curves drop one step earlier: at $\varepsilon_i=0.05$ it serves 20.2\% instead of 30.1\% in Small and none instead of 21.9\% in S2, and at $\varepsilon_i=0.001$ it serves 3.7\% instead of 6.6\% in S1.

\emph{Takeaway:} $\varepsilon_i$ gives the operator a single, exact knob between RAN reliability and AI throughput. Its cost depends on the tail of the RAN demand, which a mean and a CV do not fix; because \CAIR{} has a closed-form threshold for each demand model, the measured tail of a deployment can be used directly.

\subsection{AI Reliability Level}
\label{subsec:res-delta}

The reliability level $\delta_m$ acts on the profile sizes instead of the RAN (Fig.~\ref{fig:res-delta}). In the four-site deployment, $\delta_m=0.20$, 0.10, and 0.05 give the same plan, because the profiles chosen for throughput already keep every AI model below 1.1\% overload. At $\delta_m=0.01$, Vision at Site~3, whose \texttt{4g} profile is overloaded in 1.1\% of the intervals, would need the whole \texttt{7g} profile, which does not fit next to the RAN, so the site runs LLM7B instead and AI served falls by 29\% to 482.0 requests/s. In the larger scenarios, a looser $\delta_m$ lets the optimizer use smaller profiles: at $\delta_m=0.20$ the worst AI model is overloaded in up to 18.8\% of the intervals, while AI served rises by at most 1.3 percentage points (S3). Every plan meets its $\delta_m$.

\emph{Takeaway:} a loose AI level buys little throughput and much overload, whereas a strict one costs throughput only when a reliable profile no longer fits next to the RAN. Because AI models can move to larger profiles while the RAN demand must be covered at every site, a strict AI level is much cheaper than a strict RAN level.

\subsection{GPU Resources and Demand Variability}
\label{subsec:res-resources}

This experiment uses \CAIR{} as a planning tool: it changes the inputs that an operator controls or must estimate and asks which of them limits AI. Fig.~\ref{fig:res-resources}(a) changes the GPU resources of the four-site deployment. Adding one GPU per site leaves the RAN reservation at 33 slices, so the 28 extra slices go to AI: 16 models of all four services serve 683.3 requests/s, the full AI demand. The extra GPU also makes the strict level affordable: at $\varepsilon_i=0.01$ the RAN takes 40 slices and AI still receives 682.8 requests/s, or 99.9\% of the demand, against 460.3 requests/s on the base GPUs. On A100 40GB GPUs every profile has half the memory, and the LLM capacities are scaled by the HBM bandwidth ratio 1555/2039~\cite{NVIDIA_A100}. AI served does not change (680.8 requests/s), because every model still fits on the same slice count; only the LLM7B model at Site~4 now runs closer to its limit, with a 4.9\% overload probability.

The two CVs are the least certain inputs, so Fig.~\ref{fig:res-resources}(b) varies them. The RAN CV sets the reservation: 29, 33, and 38 slices for a CV of 0.15, 0.33, and 0.45. The AI CV sets the profile sizes: at a CV of 0.50 the reliable profiles become larger, and AI served drops to 458 to 481 requests/s whatever the RAN CV, whereas a RAN CV of 0.45 alone costs 9\%. In the larger scenarios the RAN CV matters more, because it decides which sites cross the \texttt{4g} boundary: at a CV of 0.15 Large serves 18.3\% and S2 21.9\% of the AI demand, and at 0.45 they serve 7.0\% and none. All plans in these sweeps still meet $\varepsilon_i$ and $\delta_m$.

\emph{Takeaway:} GPU slices, not memory, are the binding resource for this service mix, and one extra GPU per site removes the conflict between strict RAN protection and AI throughput. Higher variability costs throughput but not reliability, because \CAIR{} sizes every reservation and profile from the CVs. An underestimated CV would weaken the guarantees in the same way as the underestimated mean of Section~\ref{subsec:res-growth}, so the CVs should be estimated from measurements in every interval.

\subsection{Demand Levels and Service Classes in the Larger Scenarios}
\label{subsec:res-classes}

\CAIR{} is solved once per orchestration interval, so this experiment follows the larger scenarios over the four demand levels and asks what limits AI at each one. At the valley hour no RAN needs more than a \texttt{2g} instance. The smallest instance already covers most of the demand tail, so even the mean-demand plans violate the RAN in at most 5.6\% of the intervals, and both plans serve within four percentage points of each other: 34\% to 65\% of the AI demand. The limit at this hour is the number of instances rather than compute: a site offers up to 40 services, its GPU holds at most seven instances including the RAN, and an AI model serves only its own site. At the shoulder hour \CAIR{} serves 25\% to 62\% of the demand and the mean-demand plans 33\% to 64\%, but the latter violate the RAN in up to 39\% of the intervals. In the event hour \CAIR{} remains feasible in every scenario, but Large and S1 serve less than 7\% of the AI demand because 9 of their 15 RANs take the whole GPU.

The served demand is also uneven across service classes. Services with high per-slice rates are served where they are popular; S4 serves 82\% of its vision demand at the shoulder hour, and S3 serves 70\% of its NLP-encoder demand. Large LLMs with 32B to 72B parameters need large and slow profiles and receive at most 13\% of their demand in any scenario. This follows from the objective \eqref{eq:objective}, which counts the requests of all services equally.

\emph{Takeaway:} the limit on AI moves over the day, from the number of MIG instances at the valley to the RAN reservation at the busy and event hours, so a reservation sized once for the busy hour would waste capacity at night and fail at peak; \CAIR{} re-optimizes it for each interval. Which services are served follows the objective, and an operator that values LLM requests more can weight them without changing the structure of \CAIR{}.

\subsection{Unplanned Demand Growth}
\label{subsec:res-growth}

The guarantees of \CAIR{} hold for the demand distribution used for planning. This experiment measures what happens when the forecast is wrong. Fig.~\ref{fig:res-growth} keeps the busy-hour plan while both demands turn out 10\%, 20\%, or 30\% higher than forecast, and compares it with a plan re-optimized for the higher level. The worst-site RAN violation of the kept plan grows from 4.1\% to 7.7\%, 13.1\%, and 20.1\%, and every site exceeds 5\% from the first step; the worst AI overload grows from 1.1\% to 10.1\%. The re-optimized plan reserves 37, 40, and 42 RAN slices and keeps both probabilities within 5\%, at the price of AI throughput (681.4, 551.7, and 597.2 requests/s). Its AI throughput is not monotone: from $+20\%$ Site~3 keeps only three slices for AI, so its Vision model is dropped, and at $+30\%$ the same models serve more because their demand is higher.

\emph{Takeaway:} a single-stage plan is only as good as its forecast. Because \CAIR{} solves in well under a second, it can be re-solved with a fresh forecast in every interval, which restores both guarantees; extending it to adapt within an interval is the subject of the future work in Section~\ref{sec:conclusion}.

\subsection{Computation Time}
\label{subsec:res-time}

The four-site instance of \CAIR{} has 168 binary variables, 4 integer variables, and 57 constraints, and every four-site instance solves in less than 0.02~s. The largest scenario instances have about 3{,}400 binary variables and solve in at most 0.8~s.

\subsection{Discussion}
\label{subsec:res-discussion}

The experiments support each design choice of \CAIR{}. First, the chance-constrained reservation keeps the RAN within its target, whereas a mean-demand reservation violates it in 24\% to 47\% of the intervals in the four-site deployment and in 33\% to 47\% in the larger scenarios at the busy hour. Its cost to AI is small when GPUs have spare slices and large when they are busy or when the RAN sits just above a MIG instance size. Second, the reliability condition turns the expected throughput into throughput that the AI models can deliver; without it, the extra requests come from models that are overloaded most of the time. Third, multi-site replication serves up to 2.7 times more AI requests than single-site deployment, because each AI model serves only local demand. Fourth, the deterministic equivalents keep \CAIR{} exact and fast enough to be re-solved in every interval, for any of the demand models and for hardware rules such as single-instance RANs. The evaluation also exposes three limits of a static single-interval model. Its reliability holds only for the demand level used for planning; the cost of protection depends on the tail of the RAN demand, which must be measured rather than assumed; and coarse MIG instances can make a strict RAN target very expensive. These limits motivate the extensions below.
%
\section{Conclusion and Future Work}
\label{sec:conclusion}

This paper proposed \CAIR{}, a single-stage chance-constrained model that places AI models on
fixed-MIG GPUs shared with the RAN, maximizing expected AI requests served while protecting
uncertain RAN demand and enforcing AI capacity reliability. With binomial or lognormal RAN
demand and Gamma AI demand, the chance constraint and the reliability condition reduce to
thresholds computed before the solve, and \CAIR{} remains a mixed-integer linear program that
solves in less than one second for deployments with 15 sites and 40 services.

On a four-site deployment calibrated from published GPU and benchmark measurements, \CAIR{}
kept the RAN within its target at every hour, whereas a mean-demand reservation violated it in
up to 47\% of the intervals; the same held in seven larger scenarios. The price of this
protection depends on the load and on the hardware: it is negligible at off-peak hours, 4.6\%
of the AI throughput in the event hour of the four-site deployment, and all of the AI
capacity in the operator scenario, whose RANs slightly exceed the largest MIG instance below
the full GPU. The reliability condition removed AI models that would be overloaded most of the
time at peak load, and replication served up to 2.7 times more AI requests than single-site
deployment. Two further limits emerged. The guarantee holds only for the demand level used for
planning, as shown by the violations under unplanned demand growth, and the cost of strict RAN
reliability depends on the tail of the demand: with the same mean and CV, a lognormal RAN
demand needs more slices than a binomial one and cannot meet $\varepsilon_i=0.001$.

Future work will extend \CAIR{} to a two-stage stochastic formulation with recourse, such as
MIG reconfiguration and resource reassignment once the demand is observed, with explicit
treatment of reconfiguration overhead and service continuity. Because the demand of AI-native
6G networks is non-stationary, we will also study learning-based updates of the demand
distributions and reliability levels between orchestration intervals, with distributionally
robust thresholds when the distribution family is uncertain. Further extensions include
cross-site request routing, correlated RAN and AI demand, and weighted objectives that value
the requests of different services differently.
%
\appendix[Derivation of the Deterministic Equivalents]

\emph{Lognormal Threshold~\eqref{eq:dep-lognormal}.} If $\ln\widetilde d_{i,t}\sim\mathcal N(\mu_i,\sigma_i^2)$ and $\phi_i(y_i)>0$, then $\widetilde d_{i,t}\leq\phi_i(y_i)$ is the same event as $(\ln\widetilde d_{i,t}-\mu_i)/\sigma_i\leq(\ln\phi_i(y_i)-\mu_i)/\sigma_i$, whose probability is $\Phi\big((\ln\phi_i(y_i)-\mu_i)/\sigma_i\big)$. Since $\Phi$ is continuous and strictly increasing, this probability is at least $1-\varepsilon_i$ if and only if $\ln\phi_i(y_i)\geq\mu_i+\sigma_i\Phi^{-1}(1-\varepsilon_i)$, and taking the exponential of both sides gives \eqref{eq:dep-lognormal}.

For the Gamma AI demand, let $\widetilde\lambda\sim\mathrm{Gamma}(\kappa,\theta)$, with probability density and cumulative distribution functions given by
\begin{equation*}
    f_{\kappa,\theta}(\lambda)=\frac{\lambda^{\kappa-1}e^{-\lambda/\theta}}{\Gamma(\kappa)\,\theta^{\kappa}},\quad
    F_{\Gamma(\kappa,\theta)}(\mu)=\int_0^{\mu}f_{\kappa,\theta}(\lambda)\,d\lambda.
\end{equation*}

\emph{Reliability Condition~\eqref{eq:gamma-reliable}.} The condition in \eqref{eq:eligible_profiles} is $\Prb(\widetilde\lambda\leq\mu_{m,k})=F_{\Gamma(\kappa,\theta)}(\mu_{m,k})\geq1-\delta_m$. Since $F_{\Gamma(\kappa,\theta)}$ is continuous and strictly increasing on $[0,\infty)$, applying its inverse to both sides gives $\mu_{m,k}\geq F^{-1}_{\Gamma(\kappa,\theta)}(1-\delta_m)$.

\emph{Expected Number of AI Requests Served~\eqref{eq:ai-served}.} Multiplying $f_{\kappa,\theta}(\lambda)$ by $\lambda$ and using $\Gamma(\kappa+1)=\kappa\,\Gamma(\kappa)$ gives $\lambda f_{\kappa,\theta}(\lambda)=\kappa\theta\, f_{\kappa+1,\theta}(\lambda)$, since
\begin{equation*}
    \frac{\lambda^{\kappa}e^{-\lambda/\theta}}{\Gamma(\kappa)\,\theta^{\kappa}}
    =\kappa\theta\,\frac{\lambda^{(\kappa+1)-1}e^{-\lambda/\theta}}{\Gamma(\kappa+1)\,\theta^{\kappa+1}}.
\end{equation*}
Integrating from $0$ to $\mu$ yields $\E[\widetilde\lambda\,\one\{\widetilde\lambda\leq\mu\}]=\int_0^{\mu}\lambda f_{\kappa,\theta}(\lambda)\,d\lambda=\kappa\theta\,F_{\Gamma(\kappa+1,\theta)}(\mu)$.

An AI model serves the whole demand when $\widetilde\lambda\leq\mu$ and only its capacity otherwise, so $\min\{\widetilde\lambda,\mu\}=\widetilde\lambda\,\one\{\widetilde\lambda\leq\mu\}+\mu\,\one\{\widetilde\lambda>\mu\}$. Taking expectations and using $\E[\one\{\widetilde\lambda>\mu\}]=1-F_{\Gamma(\kappa,\theta)}(\mu)$,
\begin{equation*}
    \E\big[\min\{\widetilde\lambda,\mu\}\big]=\kappa\theta\,F_{\Gamma(\kappa+1,\theta)}(\mu)+\mu\big[1-F_{\Gamma(\kappa,\theta)}(\mu)\big],
\end{equation*}
which is \eqref{eq:ai-served} with $\mu=\mu_{m,k}$.

\bibliographystyle{IEEEtran}
\bibliography{references_abbreviated}

@IEEEtranBSTCTL{IEEEexample:BSTcontrol,
  CTLdash_repeated_names = "no"
}

@article{Luu2021UncertaintyAware,
  author  = {Luu, Quang-Trung and Kerboeuf, Sylvaine and Kieffer, Michel},
  title   = {Uncertainty-Aware Resource Provisioning for Network Slicing},
  journal = {IEEE Trans. Netw. Service Manag.},
  volume  = {18}, number = {1}, pages = {79--93}, month = mar, year = {2021},
  doi     = {10.1109/TNSM.2021.3058139}
}

@article{Luu2022AdmissionControl,
  author  = {Luu, Quang-Trung and Kerboeuf, Sylvaine and Kieffer, Michel},
  title   = {Admission Control and Resource Reservation for Prioritized Slice Requests With Guaranteed {SLA} Under Uncertainties},
  journal = {IEEE Trans. Netw. Service Manag.},
  volume  = {19}, number = {3}, pages = {3136--3153}, month = sep, year = {2022},
  doi     = {10.1109/TNSM.2022.3160352}
}

@article{Caballero2017SpatialLoads,
  author  = {Caballero, Pablo and Banchs, Albert and de Veciana, Gustavo and Costa-P\'erez, Xavier},
  title   = {Multi-Tenant Radio Access Network Slicing: Statistical Multiplexing of Spatial Loads},
  journal = {IEEE/ACM Trans. Netw.},
  volume  = {25}, number = {5}, pages = {3044--3058}, month = oct, year = {2017},
  doi     = {10.1109/TNET.2017.2720668}
}

@article{Shah2025CAORA,
  author  = {Shah, Syed Danial Ali and Hafeez, Maryam and Salama, Abdelaziz and Zaidi, Syed Ali Raza},
  title   = {Proactive {AI}-and-{RAN} Workload Orchestration in {O-RAN} Architectures for {6G} Networks},
  journal = {IEEE Open J. Commun. Soc.},
  volume  = {6}, pages = {7939--7954}, year = {2025},
  doi     = {10.1109/OJCOMS.2025.3608700}
}

@inproceedings{Shah2025Interplay,
  author    = {Shah, Syed Danial Ali and Nezami, Zeinab and Hafeez, Maryam and Zaidi, Syed Ali Raza},
  title     = {The Interplay of {AI}-and-{RAN}: Dynamic Resource Allocation for Converged {6G} Platform},
  booktitle = {Proc. IEEE Conf. Comput. Commun. Workshops (INFOCOM WKSHPS)},
  address   = {London, U.K.}, month = may, year = {2025},
  doi       = {10.1109/INFOCOMWKSHPS65812.2025.11152939}
}

@misc{Turkkan2024MIG,
  author       = {Turkkan, Bekir O. and Murali, Pavankumar and Harsha, Pavithra and Arora, Rohan R. and Vanloo, Gerard and Narayanaswami, Chandra},
  title        = {Optimal Workload Placement on Multi-Instance {GPUs}},
  howpublished = {arXiv:2409.06646},
  year         = {2024}
}

@article{Kundu2025AIRAN,
  author  = {Kundu, Lopamudra and Lin, Xingqin and Gadiyar, Rajesh and Lacasse, Jean-Francois and Chowdhury, Shuvo},
  title   = {{AI-RAN}: Transforming {RAN} with {AI}-Driven Computing Infrastructure},
  journal = {IEEE Commun. Mag.},
  volume  = {64}, number = {1}, pages = {168--174}, year = {2026},
  doi     = {10.1109/MCOM.001.2500018}
}

@inproceedings{LoSchiavo2024YinYangRAN,
  author    = {Lo Schiavo, Leonardo and Ayala-Romero, Jose A. and Garcia-Saavedra, Andres and Fiore, Marco and Costa-P\'erez, Xavier},
  title     = {{YinYangRAN}: Resource Multiplexing in {GPU}-Accelerated Virtualized {RANs}},
  booktitle = {Proc. IEEE Conf. Comput. Commun. (INFOCOM)},
  address   = {Vancouver, BC, Canada}, month = may,
  pages     = {721--730}, year = {2024},
  doi       = {10.1109/INFOCOM52122.2024.10621380}
}

@inproceedings{Kholmatov2025AoRA,
  author    = {Kholmatov, Siyavushkhon and Cho, Seongsik and Lee, Kyunghan and Chong, Song},
  title     = {{AoRA}: {AI}-on-{RAN} for Backhaul-Free Edge Inference},
  booktitle = {Proc. ACM SIGCOMM},
  pages     = {1263--1265}, year = {2025},
  doi       = {10.1145/3718958.3750517}
}

@article{Shah2026FederatedCAORA,
  author  = {Shah, Syed Danial Ali and Gregory, Mark A. and Hafeez, Maryam and Zaidi, Syed Ali Raza},
  title   = {Enabling Multi-Operator {AI-RAN} Coexistence Through Federated Edge Orchestration},
  journal = {IEEE Commun. Standards Mag.},
  year    = {2026}, note = {early access},
  doi     = {10.1109/MCOMSTD.2026.3690656}
}

@misc{Li2026HAF,
  author       = {Li, Haiyuan and Wu, Yulei and Simeonidou, Dimitra},
  title        = {Deadline-Driven Hierarchical Agentic Resource Sharing for {AI} Services and {RAN} Functions in {AI-RAN}},
  howpublished = {arXiv:2605.07547},
  year         = {2026}
}

@inproceedings{NeuRO2026,
  author    = {Javeed, Arshad and D\'an, Gy\"orgy and Fodor, Viktoria},
  title     = {{NeuRO}: Inference-Time Profiling and Orchestration of {ML} Applications at the Edge},
  booktitle = {Proc. IEEE Conf. Comput. Commun. (INFOCOM)},
  pages     = {1--10}, year = {2026},
  doi       = {10.1109/INFOCOM59046.2026.11571347}
}

@inproceedings{Shen2025Batching,
  author    = {Qiu, Ningyuan and Gan, Shuying and Wang, Xijun and Feng, Chenyuan and Chen, Xiang and Han, Shuangfeng and Wang, Xiaoyun},
  title     = {Joint Request Batching and Worker Assignment in {AI-RAN} Edge Inference Systems},
  booktitle = {Proc. Int. Conf. Future Commun. Netw. (FCN)},
  pages     = {1--6}, year = {2025},
  doi       = {10.1109/FCN66513.2025.11296658}
}

@inproceedings{ICUFN2026Testbed,
  author    = {Jung, Tae-il and Yoo, Hyun-Min and Kim, Kyung-Sook and Na, Jee-Hyeon and Hong, Een-Kee},
  title     = {A {5G} Testbed for {AI}-and-{RAN} on {AI} Edge: Design and Performance Evaluation},
  booktitle = {Proc. Int. Conf. Ubiquitous Future Netw. (ICUFN)},
  pages     = {589--592}, year = {2026},
  doi       = {10.1109/ICUFN69619.2026.11628529}
}

@misc{Wu2026Eunoia,
  author       = {Wu, Zongshen and Milenkovi\'c, Marko and Xing, Jiarong and Doost-Mohammady, Rahman},
  title        = {Eunoia: Toward Application-Transparent, Scalable {GPU} Sharing in {AI-RAN}},
  howpublished = {arXiv:2610.15404},
  year         = {2026}
}

@misc{Yao2026Beaver,
  author       = {Yao, Yuncheng and Qi, Zhenzhou and Zheng, Junyao and Tung, Chung-Hsuan and Zhuo, Danyang and Chen, Tingjun},
  title        = {Beaver: Elastic {GPU} Sharing Between {ML} and Latency-Critical {vRAN} Workloads},
  howpublished = {arXiv:2610.02522},
  year         = {2026}
}

@article{Feng2026AIRANPathway,
  author  = {Feng, Chenyuan and Yang, Howard H. and Guo, Kun and Xia, Wenchao and Liu, Chenxi and Quek, Tony Q. S.},
  title   = {{AI-RAN}: The Pathway to Future Wireless Networks},
  journal = {J. Inf. Intell.},
  volume  = {4}, number = {1}, pages = {5--22}, year = {2026},
  doi     = {10.1016/j.jiixd.2025.12.013}
}

@inproceedings{Li2024AIRANArchitecture,
  author    = {Li, Zexu and Wang, Qingtian and Wang, Yue and Chen, Tao},
  title     = {The Architecture of {AI} and Communication Integration Towards {6G}: An {O-RAN} Evolution},
  booktitle = {Proc. ACM MobiCom},
  pages     = {2329--2334}, year = {2024},
  doi       = {10.1145/3636534.3701550}
}

@inproceedings{Tang2025AIServiceRepo,
  author    = {Tang, Yun and Zou, Mengbang and Srinivasan, Udhaya Chandhar and Umealor, Obumneme and Kevogo, Dennis and Scott, Benjamin James and Guo, Weisi},
  title     = {Building {AI} Service Repositories for On-Demand Service Orchestration in {6G AI-RAN}},
  booktitle = {Proc. IEEE GLOBECOM},
  pages     = {4131--4136}, year = {2025},
  doi       = {10.1109/GLOBECOM59602.2025.11431783}
}

@article{Wen2026CrossDomain,
  author  = {Wen, Ruihan and Feng, Gang and Li, Chengjie and Lou, Haokang and Xu, Qiping and Liu, Tao and Chen, Zihan},
  title   = {{AI-RAN} Resource Configuration for Non-Collaborative Cross-Domain Slicing},
  journal = {J. Inf. Intell.},
  volume  = {4}, number = {1}, pages = {38--53}, year = {2026},
  doi     = {10.1016/j.jiixd.2025.12.001}
}

@article{Lee2014SpatialTraffic,
  author  = {Lee, Dongheon and Zhou, Sheng and Zhong, Xiaofeng and Niu, Zhisheng and Zhou, Xuan and Zhang, Honggang},
  title   = {Spatial Modeling of the Traffic Density in Cellular Networks},
  journal = {IEEE Wireless Commun.},
  volume  = {21}, number = {1}, pages = {80--88}, month = feb, year = {2014},
  doi     = {10.1109/MWC.2014.6757900}
}

@inproceedings{Wang2015TrafficPatterns,
  author    = {Wang, Huandong and Xu, Fengli and Li, Yong and Zhang, Pengyu and Jin, Depeng},
  title     = {Understanding Mobile Traffic Patterns of Large Scale Cellular Towers in Urban Environment},
  booktitle = {Proc. ACM Internet Meas. Conf. (IMC)},
  address   = {Tokyo, Japan}, month = oct,
  pages     = {225--238}, year = {2015},
  doi       = {10.1145/2815675.2815680}
}

@inproceedings{Li2023AlpaServe,
  author    = {Li, Zhuohan and Zheng, Lianmin and Zhong, Yinmin and Liu, Vincent and Sheng, Ying and Jin, Xin and Huang, Yanping and Chen, Zhifeng and Zhang, Hao and Gonzalez, Joseph E. and Stoica, Ion},
  title     = {{AlpaServe}: Statistical Multiplexing With Model Parallelism for Deep Learning Serving},
  booktitle = {Proc. USENIX Symp. Oper. Syst. Des. Implement. (OSDI)},
  address   = {Boston, MA, USA}, month = jul,
  pages     = {663--679}, year = {2023}
}

@inproceedings{Wang2025BurstGPT,
  author    = {Wang, Yuxin and Chen, Yuhan and Li, Zeyu and Kang, Xueze and Fang, Yuchu and Zhou, Yeju and Zheng, Yang and Tang, Zhenheng and He, Xin and Guo, Rui and Wang, Xin and Wang, Qiang and Zhou, Amelie Chi and Chu, Xiaowen},
  title     = {{BurstGPT}: A Real-World Workload Dataset to Optimize {LLM} Serving Systems},
  booktitle = {Proc. ACM SIGKDD Conf. Knowl. Discovery Data Mining (KDD)},
  address   = {Toronto, ON, Canada}, month = aug,
  pages     = {5831--5841}, year = {2025},
  doi       = {10.1145/3711896.3737413}
}

@inproceedings{Xiang2026ServeGen,
  author    = {Xiang, Yuxing and Li, Xue and Qian, Kun and Zhang, Yan and Yu, Wenyuan and Zhai, Ennan and Jin, Xin and Zhou, Jingren},
  title     = {{ServeGen}: Workload Characterization and Generation of Large Language Model Serving in Production},
  booktitle = {Proc. USENIX NSDI},
  pages     = {1845--1859}, year = {2026}
}

@inproceedings{Reddi2020MLPerf,
  author    = {Reddi, Vijay Janapa and others},
  title     = {{MLPerf} Inference Benchmark},
  booktitle = {Proc. ACM/IEEE Int. Symp. Comput. Archit. (ISCA)},
  address   = {Valencia, Spain}, month = {May/Jun.},
  pages     = {446--459}, year = {2020},
  doi       = {10.1109/ISCA45697.2020.00045}
}

@misc{MLPerfV30,
  author = {{MLCommons}},
  title  = {{MLPerf Inference} v3.0 Results, Closed Division, {NVIDIA DGX-A100} ({A100-SXM-80GB} $\times$8), Server Scenario},
  year   = {2023},
  url    = {https://github.com/mlcommons/inference_results_v3.0},
  note   = {{Accessed}: Oct. 28, 2026}
}

@misc{MLPerfV31,
  author = {{MLCommons}},
  title  = {{MLPerf Inference} v3.1 Results, Closed Division, {Dell XE9680} ({A100-SXM-80GB} $\times$8), Server Scenario},
  year   = {2023},
  url    = {https://github.com/mlcommons/inference_results_v3.1}
}

@misc{NVIDIA_MIG,
  author = {{NVIDIA}},
  title  = {{NVIDIA Multi-Instance GPU} User Guide},
  year   = {2026},
  url    = {https://docs.nvidia.com/datacenter/tesla/mig-user-guide/},
  note   = {{Accessed}: Oct. 28, 2026}
}

@misc{NVIDIA_A100,
  author       = {{NVIDIA}},
  title        = {{NVIDIA A100 Tensor Core GPU} Architecture},
  howpublished = {White paper},
  year         = {2020},
  url          = {https://images.nvidia.com/aem-dam/en-zz/Solutions/data-center/nvidia-ampere-architecture-whitepaper.pdf},
  note   = {{Accessed}: Oct. 28, 2026}
}

@misc{NVIDIA_Aerial231,
  author = {{NVIDIA}},
  title  = {{Aerial SDK} 23-1 Product Brief},
  year   = {2023},
  url    = {https://docs.nvidia.com/aerial/archive/aerial-sdk/23-1/text/product_brief/product_brief_intro.html},
  note   = {{Accessed}: Oct. 28, 2026}
}

@misc{Touvron2023Llama2,
  author       = {Touvron, Hugo and others},
  title        = {{Llama 2}: Open Foundation and Fine-Tuned Chat Models},
  howpublished = {arXiv:2307.09288},
  year         = {2023}
}

@misc{IBM_CPLEX,
  author = {{IBM}},
  title  = {{ILOG CPLEX Optimization Studio} 22.2},
  year   = {2026},
  url    = {https://www.ibm.com/products/ilog-cplex-optimization-studio},
  note   = {{Accessed}: Oct. 28, 2026}
}

@article{Huangfu2018HiGHS,
  author  = {Huangfu, Qi and Hall, J. A. Julian},
  title   = {Parallelizing the Dual Revised Simplex Method},
  journal = {Math. Program. Comput.},
  volume  = {10}, number = {1}, pages = {119--142}, year = {2018},
  doi     = {10.1007/s12532-017-0130-5}
}

@misc{SoftBank2024GH200,
  author = {{SoftBank Corp.}},
  title  = {{SoftBank Corp.} Develops {5G L1} Software, Achieving Carrier-Grade High Performance and Quality {vRAN} on {NVIDIA Grace Hopper} Platform},
  howpublished = {Press release},
  month  = nov, year = {2024},
  url    = {https://www.softbank.jp/en/corp/news/press/sbkk/2024/20241113_03/}
}

@book{Casella2002Inference,
  author    = {Casella, George and Berger, Roger L.},
  title     = {Statistical Inference},
  edition   = {2nd},
  publisher = {Duxbury},
  address   = {Pacific Grove, CA, USA},
  year      = {2002}
}

\end{document}